\documentclass[%
reprint,
bibnotes, amsmath, amssymb, aps, pre,
 showkeys,
floatfix
]{revtex4-2}

\usepackage{graphicx}
\usepackage[section]{placeins}
\usepackage{float}

\usepackage{dcolumn}
\usepackage{multirow}
\usepackage{bm}
\usepackage{bbm}
\usepackage[colorlinks=true,linkcolor=blue,citecolor=red]{hyperref}
\usepackage{physics}
\usepackage{nicefrac}
\usepackage{siunitx}
\usepackage{array}

\usepackage[normalem]{ulem}
\usepackage{cancel}
\newcommand{\stkout}[1]{\ifmmode\text{\sout{\ensuremath{#1}}}\else\sout{#1}\fi}

\newcommand{\kb}{k_{\mathrm{B}}}
\newcommand{\kT}{k_{\mathrm{B}}T}

\newcommand{\mrm}{\mathrm}
\newcommand{\mcal}{\mathcal}

\newcommand{\pr}[1]{\left(#1\right)} 
\newcommand{\sr}[1]{\left[#1\right]} 
\newcommand{\br}[1]{\left\{#1\right\}} 

\newcommand{\trelo}{\tau_{\mrm{R}}}

\newcommand{\dg}{\delta_{\mrm{g}}}
\newcommand{\ts}{t_{\mrm{s}}}
\newcommand{\fs}{f_{\mrm{s}}}

\newcommand{\dlk}{\Delta\lambda_{k}}

\newcommand{\xk}{x_{k}}
\newcommand{\xkp}{x_{k+1}}

\newcommand{\lk}{\lambda_{k}}
\newcommand{\lkp}{\lambda_{k+1}}

\newcommand{\xkr}{r_{k}}
\newcommand{\xkrp}{r_{k+1}}

\newcommand{\xkpr}{r_{k}^{+}}

\newcommand{\xkrt}{\Delta x_{k}^{\mrm{tot}}}
\newcommand{\xkprt}{\Delta x_{k^{+}}^{\mrm{tot}}}

\newcommand{\wt}{W^{\mrm{trap}}}
\newcommand{\wg}{W^{\mrm{grav}}}
\newcommand{\wn}{W^{\mrm{net}}}
\newcommand{\pt}{P^{\mrm{trap}}}
\newcommand{\pg}{\dot{F}}
\newcommand{\pn}{P^{\mrm{net}}}

\newcommand{\ept}{\ev{P^{\mrm{trap}}}}
\newcommand{\epg}{\langle\dot{F}\rangle}
\newcommand{\epn}{\ev{P^{\mrm{net}}}}
\newcommand{\xT}{X_{\mrm{T}}}
\newcommand{\xR}{X_{\mrm{R}}}
\newcommand{\eptm}{\overline{\ev{P^{\mrm{trap}}}}}

\newcommand{\mP}{\mcal{P}}

\newcommand{\tmfp}{\ev{\tau_{\mrm{FP}}}}

\DeclareMathOperator{\e}{e}
\DeclareMathOperator*{\argmax}{arg\,max}

\begin{document}

\title{Maximal Fluctuation Exploitation in Gaussian Information Engines}

\author{Joseph N.\ E.\ Lucero}
\altaffiliation{Current address: Department of Chemistry, Stanford University, Stanford, CA, 94305 USA}
\author{Jannik Ehrich}
\author{John Bechhoefer}
\author{David A.\ Sivak}
\email{dsivak@sfu.ca}
\affiliation{Department of Physics, Simon Fraser University, Burnaby, BC, V5A 1S6 Canada}%

\date{\today}

\begin{abstract}
Understanding the connections between information and thermodynamics has been among the most visible applications of stochastic thermodynamics.
While recent theoretical advances have established that the second law of thermodynamics sets limits on information-to-energy conversion, it is currently unclear to what extent real systems can achieve the predicted theoretical limits.
Using a simple model of an information engine that has recently been experimentally implemented, we explore the limits of information-to-energy conversion when an information engine's benefit is limited to output energy that can be stored.
We find that restricting the engine's output in this way can limit its ability to convert information to energy. Nevertheless, a feedback control that inputs work can allow the engine to store energy at the highest achievable rate.
These results sharpen our theoretical understanding of the limits of real systems that convert information to energy.
\end{abstract}

\maketitle

\section{Introduction}

At the dawn of statistical mechanics 150 years ago, Maxwell proposed a thought experiment that has come to be known as ``Maxwell's demon,'' a device that harnesses thermal fluctuations to extract work seemingly without incurring any of the dissipative costs mandated by the second law of thermodynamics~\cite{Maxwell1878}. 
This thought experiment challenged our understanding of the role observations and information play in thermodynamics.

Seventy years later, Leo Szilard proposed a simpler implementation of Maxwell's demon consisting of a single particle in a box, connected to a thermal reservoir at temperature $T$~\cite{Szilard1964}.
A partition is inserted into the middle of the box, and the demon measures the location of the particle within the box, thereby collecting information about the particle.
A weight is then appropriately attached to the partition via a pulley.
By letting the gas slowly expand to its original volume, it exerts a force against the partition, extracting $\kT\log 2$ of useful work. 
Thus, the information collected about the particle is used to extract work via an isothermal process. 
Such a device functions as an \emph{information engine}~\cite{Paneru2020}. 
Later, it was realized that this apparent violation of the second law can be resolved by recognizing the unavoidable cost of the information processing~\cite{leff_book2002,Bennett1982}. 

Since the first formulation of Maxwell's demon, physicists and engineers have sought to create experimental realizations. 
Many of the proposed designs have been impractical because rectifying thermal fluctuations requires resolving system dynamics on small length and time scales.
Nonetheless, successful implementations have recently been demonstrated~\cite{Toyabe2010,Lee2018,admon2018,Ribezzi-Crivellari2019,Paneru2018b} and have been used to probe the estimated thermodynamic cost of information processing and the efficiency of information-to-work conversion. 
These technological advances, aided by simultaneous advances in our theoretical understanding of stochastic thermodynamics~\cite{Cao2009,Jarzynski2011,seifert_2012}, have stimulated interest in exploring the connections between information and thermodynamics~\cite{Parrondo2015a}. 

Recently, we experimentally realized an information engine based on a colloidal particle, which successfully extracts energy from a thermal reservoir and stores this energy by raising a weight against gravity~\cite{Saha2021}.
We systematically optimized this information engine under the constraint that no external work be done on the particle, finding that there is an optimal particle mass that maximizes the energy-storage rate (the output power).

In this article, we generalize the design of our engine by relaxing the no-work constraint and explore the broader range of its performance---as quantified by the net rate of energy extraction (net output power)---made possible by allowing the input of external work.
We use multi-objective optimization to identify feedback rules that maximize the trade-off between output and input of the information engine.
We elucidate the physical mechanism underlying these optimal feedback rules; we further compare the performance of information engines that can extract and store all of the particle's energy to those engines whose output is restricted to only stored free-energy.

We find that there is an upper bound to the rate of energy extraction, even when information about the particle's position is collected continuously and with perfect accuracy, and all of the particle's potential energy can be stored.
We further show that restricting the information engine's output to stored free energy limits the maximal rate of net energy extraction.
However, for sufficiently heavy particles, we show that an appropriate feedback-control rule ensures the maximal rate. 
We find that these feedback rules use work input into the system to reduce the duration of unproductive excursions, thereby enhancing the overall output power of the engine.

The paper is organized as follows: 
In Sec.~\ref{sec:model}, we lay out the model, detailing the potential energy and corresponding equilibrium distribution in Sec.~\ref{subsec:potential_energy_equil} and the dynamical equations of motion in Sec.~\ref{subsec:equations_of_motion}.
We establish the thermodynamics of the information engine in Sec.~\ref{subsec:engine_thermo} and the associated constraints in Sec.~\ref{subsec:accounting_considerations}.
Section~\ref{subsec:performance_metrics} introduces the measure of information-engine performance that we optimize under various constraints in Sec.~\ref{sec:results}.
Specifically, Sec.~\ref{sec:full_extraction_scheme} examines the case where energy can be extracted both as a free-energy change and as work from the trap; 
Sec. \ref{subsec:practical_storage_schemes} investigates the case where the engine is restricted to extract energy only as a free-energy change.
Section~\ref{sec:conclusion} summarizes the results and provides future outlooks for this information engine.

\section{Model}
\label{sec:model}

\subsection{Potential energy and equilibrium distribution}
\label{subsec:potential_energy_equil}

\begin{figure}[htbp]
    \centering
    \includegraphics[trim = 1 1 1 1, clip, width=\linewidth]{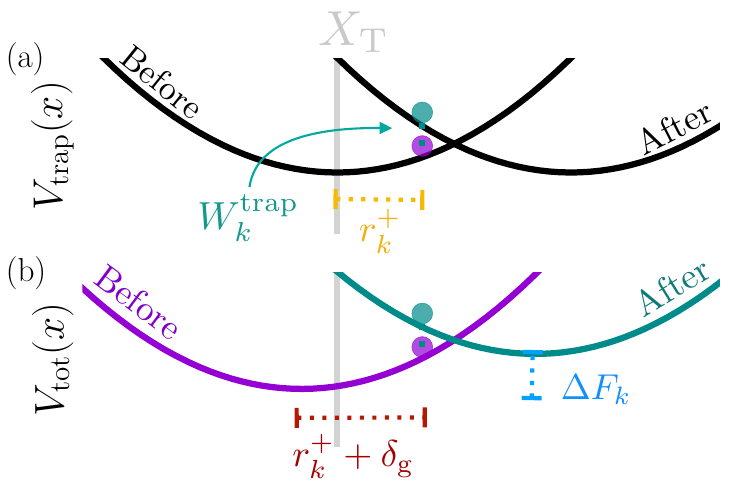}
    \caption{
        Schematic of information-engine operation in the lab frame.
        (a) The particle in the trap potential before and after a ratchet event. 
        (b) The particle in the corresponding total (trap plus gravity) potential before (purple curve) and after (teal curve) a ratchet event.
        The particle evolves under the total potential with same color.
        }
    \label{fig:lab_ratchet_schematic}
\end{figure}

As a model for an information engine, we use the setup from our previous work~\cite{Saha2021}, where a particle with effective mass $m$ immersed in a fluid medium diffuses while under the influence of gravity and a harmonic potential (Fig.~\ref{fig:lab_ratchet_schematic}), henceforth the \emph{trap} potential. 
The particle experiences a total potential $V^{\prime}$ that is the sum of the trap and gravitational potentials,
\begin{align}
    V^{\prime}(x^{\prime},\lambda^{\prime}) = \underbrace{\tfrac{1}{2}\kappa\pr{x^{\prime}-\lambda^{\prime}}^{2}}_{\text{trap}} 
            + \underbrace{mgx^{\prime}}_{\text{gravity}},\label{eq:pot_w_units}
\end{align}
where $x^{\prime}$ is the particle position (in dimensional units), $\lambda^{\prime}$ is the position of the minimum of the trap potential, $\kappa$ is the curvature (stiffness) of the trap potential, and $g$ is the acceleration due to gravity. 
The particle's effective mass $m \equiv (4/3)\pi r^{3}\Delta\rho$ depends on the relative density $\Delta\rho \equiv \rho_{\mrm{particle}} - \rho_{\mrm{medium}}$ 
and accounts for buoyancy.
For a static potential, the system evolves to a Gibbs-Boltzmann equilibrium distribution of position $x^\prime$ that is a Gaussian, $\mcal{N}(x^\prime;\mu, \sigma^{2})$, with mean $\mu = \lambda'-mg/\kappa$ and variance $\sigma^{2} = \kT/\kappa$:
\begin{align}
    \pi^{\mrm{eq}}\pr{x^{\prime}|\lambda^{\prime}} \sim \mcal{N}\pr{x^{\prime}; \lambda^{\prime}-\frac{mg}{\kappa},\frac{\kT}{\kappa}}.
\end{align}
Here, $\kb$ is the Boltzmann constant, and $T$ is the temperature of the surrounding thermal reservoir. 
The standard deviation $\sigma$ of this distribution defines a natural length scale, while the curvature $\kappa$ of the harmonic potential and the friction coefficient $\gamma$ together determine the relaxation time $\trelo\equiv\gamma/\kappa$ within the trap potential.  
To simplify expressions, we rescale all times by $\trelo$, all lengths by $\sigma$, and all energies by $\kT$. 
With this scaling, the equilibrium distribution simplifies to
\begin{align}
    \pi^{\mrm{eq}}\pr{x|\lambda} \sim \mcal{N}\pr{x; \lambda-\dg, 1}\ , \label{eq:eq_dist}
\end{align}
with $\dg \equiv mg/(\kappa\sigma)$ the scaled effective mass. 
Note that $\dg=1$ corresponds to an effective mass that in equilibrium under gravity sags a distance $\sigma$.
The corresponding scaled quantities are denoted without primes.

\subsection{Equations of motion}
\label{subsec:equations_of_motion}
The dynamics of the particle within the trap is described by the stochastic equation of motion
\begin{align}
    \dot{x} = -\pr{x-\lambda} - \dg + \sqrt{2}\ \eta(t)\ .\label{eq:od_langevin}
\end{align}
Here, $\eta(t)$ is a Gaussian white noise with zero mean and covariance $\ev{\eta(t)\eta(s)} = \delta(t-s)$, and dots above a variable denotes a time derivative.

We measure the particle's position periodically at frequency $\fs$ (or equivalently with period $\ts = 1/\fs$). 
Integrating \eqref{eq:od_langevin} from time $t$ to time $t+\ts$ gives the discrete-time stochastic equation of motion~\cite[Ch.~4.5.4]{gardiner_book2009} for timestep $k$, where $x_{k} \equiv x(k\ts)$ and $\lk \equiv \lambda\pr{k\ts}$:
\begin{align}
    \xkp = \xk\e^{-\ts} + \pr{1-\e^{-\ts}}\pr{\lk-\dg} + \sigma_{\ts}\, \xi_{k}\ . \label{eq:discrete_od_langevin}
\end{align}
The $\xi_{k}$ are independent Gaussian random variables with zero mean and unit variance, and $\sigma_{\ts}^{2} = 1-\e^{-2\ts}$ is the sampling-period-dependent variance.  

In response to this measurement, we instantaneously move the trap center an increment $\dlk$~\footnote{In experimental realizations, there is typically feedback latency -- a delay between the measurement time and the response time. 
This arises from the time required to transfer the measurement to the core processor of the hardware, compute the response, and communicate to the device that moves the trap. A typical feedback latency is $\ts$, which leads to small performance reductions that are neglected here.}:
\begin{align}
    \lkp = \lk + \dlk\ . \label{eq:trap_update}
\end{align}

\begin{figure}[htbp]
    \centering
    \includegraphics[trim = 1 1 1 1, clip, width=\linewidth]{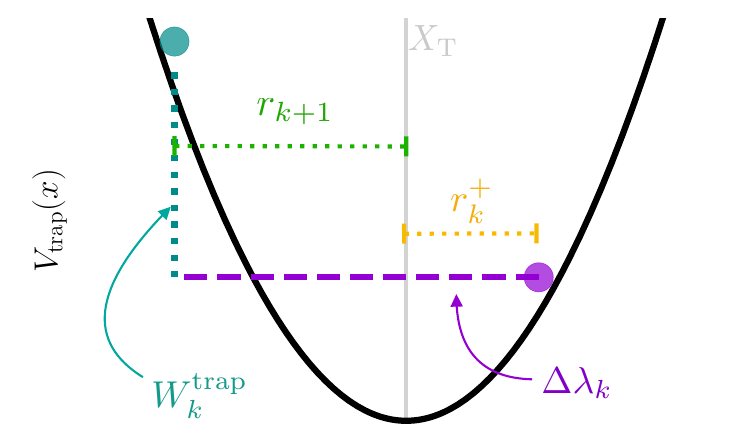}
    \caption{
        Schematic of information-engine operation in the trap frame.
        In this frame, a ratchet event (as dictated by the feedback rule) instantaneously transports the particle from its measured position $\xkpr$ (purple) to another position (teal) $\dlk$ away (dashed line), inputting work $\wt_{k}$ (vertical dotted line).
        }
    \label{fig:relative_ratchet_schematic}
\end{figure}

For theoretical convenience, we consider the particle dynamics in the relative frame of the trap potential (Fig.~\ref{fig:relative_ratchet_schematic}), henceforth the \emph{trap frame}.
The relevant degrees of freedom in this reference frame are the relative particle position immediately after measurement, $\xkpr \equiv \xkp-\lk$, and immediately following the response to the measurement, $\xkrp \equiv \xkp-\lkp$. 
In the trap frame, the discrete-time equations of motion~\eqref{eq:discrete_od_langevin} and \eqref{eq:trap_update} are 
\begin{subequations}\label{eq:relative_update}
    \begin{align}
        \xkpr &= \xkr\e^{-\ts} - \pr{1-\e^{-\ts}}\dg + \sigma_{\ts}\xi_{k}\label{eq:relative_pos_update}\\
        \xkrp &= \xkpr - \dlk \ . \label{eq:relative_trap_update}
    \end{align}
\end{subequations}
In this frame, following measurement of particle position~\eqref{eq:relative_pos_update}, the feedback shifts the particle position according to \eqref{eq:relative_trap_update}.

We consider feedback that takes the general form
\begin{align}
    \dlk &= \Theta\pr{\xkpr-\xT}\pr{\alpha\xkpr + \psi}\ , \label{eq:gen_scheme} 
\end{align}
which is a function of the current measured particle displacement $\xkpr$ from the trap center.
The increments $\dlk$ are parameterized by the proportionality $\alpha$, the offset $\psi$, and the threshold $\xT$, as measured from the trap center. 
The Heaviside step function $\Theta\pr{\cdot}$ implements the ratchet threshold condition.
Specification of these parameters $\{\alpha,\psi,\xT\}$ constitutes a \emph{feedback rule}.
We say that a \emph{ratchet event} occurs when evaluation of \eqref{eq:gen_scheme} produces an increment $\dlk\neq 0$.

The feedback implies that if the relative particle position from the trap center exceeds the threshold $\xT$, then the particle position in the trap frame (or the trap center in the lab frame) is shifted in response; otherwise, the position is unchanged.
The amount moved during a ratchet event is a constant plus a term proportional to the displacement (i.e., an affine transformation of the displacement).

\subsection{Engine thermodynamics}
\label{subsec:engine_thermo}

Measurement of the particle's displacement leads to a feedback response~\eqref{eq:gen_scheme} that causes the trap to do work
\begin{align}
    \wt_{k} = \frac{1}{2}\sr{\pr{\xkrp}^{2}-\pr{\xkpr}^{2}} \label{eq:trap_work_def}
\end{align}
on the particle (Fig.~\ref{fig:lab_ratchet_schematic}a; Fig.~\ref{fig:relative_ratchet_schematic}).
By convention, this work is negative when the particle's potential energy decreases, in which case energy is extracted from the particle as work via the trap.

The particle's equilibrium free energy is 
\begin{align}
    F \equiv 
        \ev{ 
            V(x,\lambda) + \ln \pi^{\mrm{eq}} ( x | \lambda ) 
        }_{\pi^{\mrm{eq}}(x|\lambda)} \ .
\label{eq:free_energy}
\end{align}
Upon applying feedback, this free energy changes by~\footnote{In general, the particle will be in a nonequilibrium state; however, we consider the equilibrium free energy since we implicitly assume that the particle relaxes to equilibrium at the end of the process. 
Therefore, we can understand the gain in gravitational potential energy through the feedback operation as a change in free energy which can be related to the work input. 
As the equilibrium distribution is a function only of the difference $x-\lambda$, the equilibrium distribution at the end of the process (which the system would eventually relax to if nothing else were to happen) would be unchanged.
As a result, the free energy changes according to \eqref{eq:free_energy_change}.}
\begin{align}
    \Delta F_{k} = \delta_g \, (\lambda_{k+1} - \lambda_k)\ . \label{eq:free_energy_change}
\end{align}
The particle then diffuses for a time $\ts$ until the next measurement.  
The rate of free-energy change, averaged over all steps of the protocol, is 
\begin{subequations}
    \begin{align}
        \epg &= \frac{1}{t_{\mrm{prot}}}\sum_{k=1}^{N_{\mrm{meas}}} \Delta F_k\\
        &= \frac{1}{N_{\mrm{meas}}}\sum_{k=1}^{N_{\mrm{meas}}}\frac{\Delta F_k}{\ts}\ , \label{eq:nat_pgrav_def}
    \end{align}
\end{subequations}
where $N_{\mrm{meas}}~\equiv~\lfloor t_{\mrm{prot}}/\ts \rfloor$ is the number of measurements made over protocol duration $t_{\mrm{prot}}$ for sampling period $\ts$.
We generally consider cases where $t_{\mrm{prot}} \gg \ts$ resulting in many measurements for each trajectory.
Similarly, the trap does work on the system at average rate
\begin{align}
    \ept &= \frac{1}{N_{\mrm{meas}}}\sum_{k=1}^{N_{\mrm{meas}}}\frac{\wt_{k}}{\ts} \ . \label{eq:nat_ptrap_def}
\end{align}
We then define the \textit{net output power} as the difference between the rate of free-energy change and the input trap work,
\begin{subequations}
    \begin{align}
        \epn &= \epg - \ept \\
        &= \frac{1}{N_{\mrm{meas}}}\sum_{k=1}^{N_{\mrm{meas}}}\frac{\Delta F_{k}-\wt_{k}}{\ts} \ .\label{eq:nat_pnet_def}
    \end{align}
\end{subequations}

\subsection{Accounting considerations}
\label{subsec:accounting_considerations}

In general, measurement-feedback processes allow the extraction of work from an isothermal reservoir~\cite{Parrondo2015a}. 
The conventional second law must be modified to explicitly include the information gathered during the measurement~\cite{Sagawa2010,Horowitz2010,Sagawa2012}, reflecting minimal costs of operating the measuring device.
When these costs are not considered, this leads to apparent violations of the second law,
\begin{align}
    \left\langle W \right\rangle - \Delta F < 0\ ,
\end{align}
including extraction of work at constant free energy ($\ev{W}<0$ and $\Delta F =0$), increased free energy without any work ($\ev{W}=0$ and $\Delta F>0$), and combinations thereof.

\begin{figure}[htbp]
    \centering
    \includegraphics[trim = 1 1 1 1, clip, width=\linewidth]{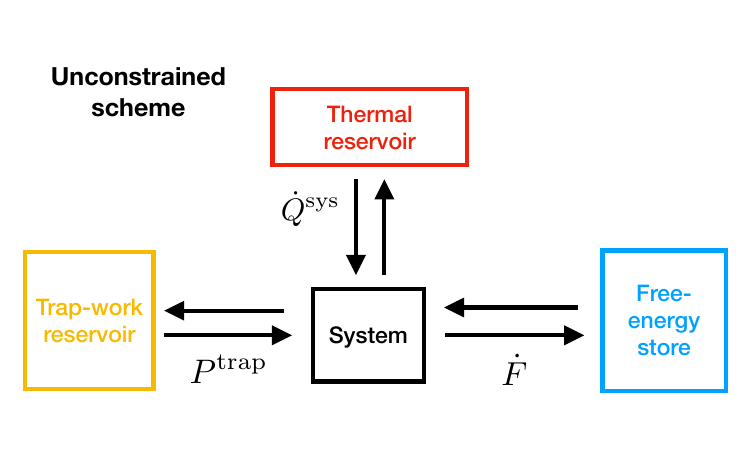}
    \caption{
        Energy flows among the system, free-energy store, trap-work reservoir, and thermal reservoir, for the unconstrained scheme. 
        }
    \label{fig:full_extraction_diagram}
\end{figure}

The \emph{unconstrained} feedback scheme~\eqref{eq:gen_scheme} permits positive or negative increments for free energy~\eqref{eq:nat_pgrav_def} and trap work~\eqref{eq:nat_ptrap_def} and thus bidirectional energy exchange among the system, the free-energy store, and the trap-work reservoir (Fig.~\ref{fig:full_extraction_diagram}). 
As such, this feedback scheme allows energy to be stored as work via the trap potential or as free energy in the gravitational potential.

Although the unconstrained feedback scheme does not present a conceptual challenge, there are practical experimental difficulties in storing work via the trap (i.e., negative trap work).
Information engines have so far been able to extract and store energy via a flow field~\cite{admon2018} or gravity~\cite{Saha2021}.
However, all other energy flows not directly linked to either an increase in the system's internal energy via the trap or an increase in system free energy are quickly lost to the environment as heat. 
We refer to feedback schemes whose output is limited to stored free energy as \emph{practical-storage} feedback schemes.

Our setup also restricts the free-energy change.
The free energy in \eqref{eq:free_energy} has two contributions: an \emph{energetic} one through the equilibrium potential energy $\ev{V(x,\lambda)}_{\pi^{\mrm{eq}}(x|\lambda)}$ and an \emph{entropic} one through $-\ev{\ln \pi^{\mrm{eq}}(x|\lambda)}_{\pi^{\mrm{eq}}(x|\lambda)}$. 
Measurement and feedback can in principle influence both contributions; however, we choose to only change the potential energy by moving the trap ``up'' or ``down'', hence \eqref{eq:free_energy_change} is defined entirely in terms of changes in the trap minimum location. 
If one did increase the trap stiffness, the system's entropy would decrease, increasing the free energy in response to a measurement~\cite{Granger2016, Dinis2021}; here, we focus on free-energy storage via the gravitational potential.

\begin{figure}[htbp]
    \centering
    \includegraphics[trim = 1 1 1 1, clip, width=\linewidth]{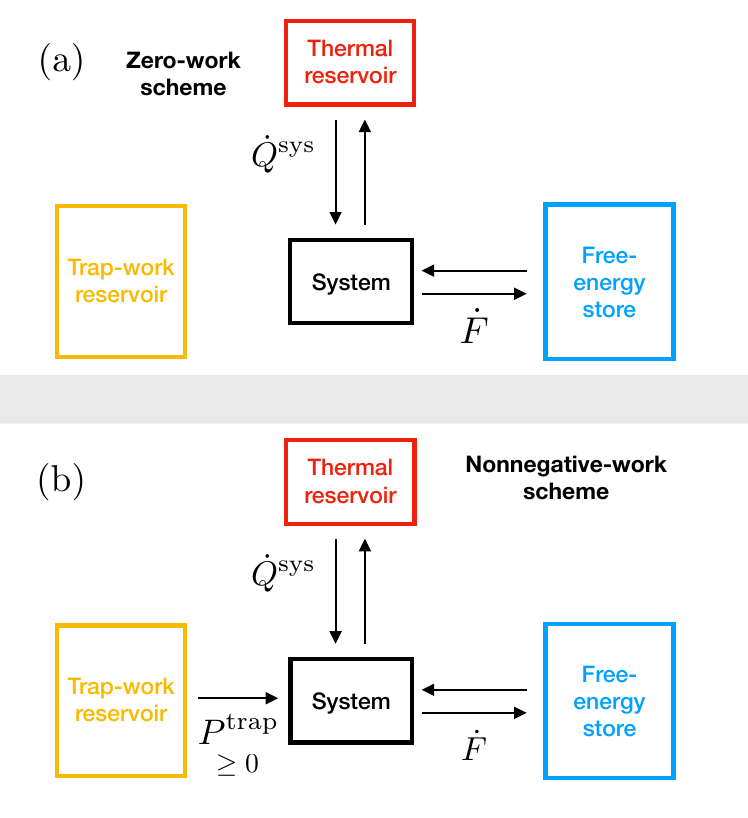}
    \caption{ 
        Energy flows among the system, the free-energy store, trap-work reservoir, and the thermal reservoir, for the (a) zero-work and (b) nonnegative-work feedback schemes. 
        }
    \label{fig:practical_extraction_diagram}
\end{figure}

We focus on two ways to model the practical constraints that restrict the output to only equilibrium free-energy changes (see App.~\ref{sec:diss_feedback} for a third way).
One method completely eliminates the trap-work reservoir, by forbidding any trap work during any ratchet event via the feedback scheme
\begin{align}
    \dlk &= \Theta\pr{\xkpr-\xT}2\xkpr \ , \label{eq:pnas_rule}
\end{align}
i.e., \eqref{eq:gen_scheme} with $\alpha=2$ and $\psi=0$.
The threshold $\xT$ is chosen to maximize the net output power.
This scheme uses only information (and no trap work) to lift the particle and store free energy [Fig.~\ref{fig:practical_extraction_diagram}a].
We call this the \emph{zero-work} feedback scheme. 
It was explored extensively in our previous work~\cite{Saha2021} and is discussed in Sec.~\ref{subsubsec:maxwell_demon_feedback_scheme}.

Another way to implement practical considerations is to design a scheme in which the allowed increments $\dlk$ are only those that produce non-negative trap work ($\wt_{k} \ge 0\ \forall\ k$):
\begin{subequations}\label{eq:strict_scheme}
    \begin{align}
        \dlk^{\mrm{prop}} &= \Theta\pr{\xkpr-\xT}\pr{\alpha\xkpr + \psi}\label{eq:strict_scheme1}\\
        \dlk &= \Theta\sr{\wt\pr{\dlk^{\mrm{prop}}}}\dlk^{\mrm{prop}} \ .\label{eq:strict_scheme2}
    \end{align}
\end{subequations}
This feedback scheme proposes an increment $\dlk^{\mrm{prop}}$ based on the most recent measurement of the particle's position and rejects ratchet events with negative work, corresponding to energy extracted from the particle.
As such, we call this the \emph{nonnegative-work} feedback scheme.
By construction, this scheme enforces unidirectional energy flow from the trap into the system [Fig.~\ref{fig:practical_extraction_diagram}b].

\subsection{Performance metrics}
\label{subsec:performance_metrics}

Since the rate $\pg$ of free-energy change and the trap power $\pt$ cannot in general be simultaneously optimized, we examine the Pareto frontier, the set of Pareto-optimal rules that are not dominated by any other rule (hence cannot improve one objective without worsening another objective)~\cite{Solon2018,Seoane2016}.
This defines the limits of (suitably defined) performance.
To obtain the Pareto frontier for the considered feedback schemes, we vary the parameters $\{\alpha, \psi, \xT\}$ to maximize the single-objective performance function
\begin{align}
    \mP(\Gamma) \equiv 2\Gamma\epg - 2\pr{1-\Gamma}\ept \ , \label{eq:gen_ratchet_benefit}
\end{align}
given a value of the trade-off parameter $\Gamma\in(0,1)$.
Shifting $\Gamma$ from 0 to 1 shifts from solely minimizing the cost (input trap work) to solely maximizing the gain (free-energy extraction).
At intermediate $\Gamma$, the rate $\epg$ of free-energy change and the input trap power $\ept$ face a finite trade-off. 
We include factors of 2 on the right-hand side of \eqref{eq:gen_ratchet_benefit} to ensure that $\Gamma=1/2$ equates the performance with the net output power~\eqref{eq:nat_pnet_def}: $\mP\pr{\Gamma=1/2} = \epn$.

\section{Results}
\label{sec:results}

\subsection{Unconstrained feedback scheme}
\label{sec:full_extraction_scheme}
We first optimize the unconstrained feedback scheme by varying the feedback parameters.
Given that this scheme can store all the potential energy from the particle, either as work via the trap or as a free-energy change via gravitational potential, it stands to reason that it would be suboptimal if the scheme were to take advantage only of a select set of fluctuations.
We therefore hypothesize that the optimal feedback scheme does not have a finite threshold, so that a ratchet event follows every measurement, simplifying \eqref{eq:gen_scheme} to
\begin{align}
    \dlk &= \alpha\xkpr + \psi\ .\label{eq:natural_ansatz}
\end{align}
We refer to this special case of the unconstrained scheme as the \emph{no-threshold} feedback scheme.

\subsubsection{No-threshold steady-state derivation}

Since both the diffusive particle dynamics and the feedback rule are linear, the relative positions~\eqref{eq:relative_update} are both Gaussian processes. 
Consequently, the steady-state distributions arising from \eqref{eq:natural_ansatz} are also Gaussian. 
Deriving self-consistent equations (see App.~\ref{sec:ss_deriv} for details), we find that the steady-state distribution for the relative coordinate $\xkpr$ immediately after measurement is 
\begin{subequations}\label{eq:ss_natural_dxkt}
    \begin{align}
        \pi\pr{\xkpr} &= \mcal{N}\pr{\xkpr; \mu_{\xkpr}, \sigma_{\xkpr}^{2}} \\[3pt]
        \mu_{\xkpr} &=  -\frac{\dg + (\psi-\dg)\e^{-\ts}}{1+(\alpha-1)\e^{-\ts}} \\[3pt] 
        \sigma_{\xkpr}^{2} &= \frac{1-\e^{-2\ts}}{1-(\alpha-1)^{2}\e^{-2\ts}}\ ,
    \end{align}
\end{subequations}
while the steady-state distribution of the relative coordinate $\xkr$ immediately after feedback is
\begin{subequations}\label{eq:ss_natural_xkr}
    \begin{align}
        \tilde{\pi}\pr{\xkr} &= \mcal{N}\pr{\xkr; \mu_{\xkr}, \sigma_{\xkr}^{2}}\\[3pt]
        \mu_{\xkr} &= \frac{\pr{1-\e^{-\ts}}(\alpha-1)\dg-\psi}{1+(\alpha-1)\e^{-\ts}}\\[3pt]
        \sigma_{\xkr}^{2} &= \frac{\pr{1-\e^{-2\ts}} (\alpha-1)^{2}}{1-(\alpha-1)^{2}\e^{-2\ts}}\ .
    \end{align}
\end{subequations}
The variances of both these distributions must remain non-negative, constraining $\alpha$ to 
\begin{align}
    1-\e^{\ts} < \alpha < 1+\e^{\ts} \ .\label{eq:alpha_constraint}
\end{align}
Outside this range, the system is unstable, never settling down into a steady-state distribution.

Since at steady state $\mu_{\xkrp}=\mu_{\xkr}$, the rate of free-energy change~\eqref{eq:free_energy_change} is proportional to the difference of the means of the steady-state distributions, 
\begin{align}
    \epg = \frac{\dg}{\ts}\pr{\mu_{\xkpr}-\mu_{\xkr}} \ .\label{eq:grav_astz_power_rel_coords}
\end{align}
Similarly, the trap power~\eqref{eq:trap_work_def} is proportional to the difference of the second moments \footnote{Here, we have used that at steady state $\ev{\pr{\xkrp}^{2}} = \ev{\pr{\xkr}^{2}}$.}
\begin{subequations}\label{eq:trap_astz_power_rel_coords}
    \begin{align}
        \ept &= \frac{1}{2\ts}\sr{\ev{\pr{\xkr}^{2}}-\ev{\pr{\xkpr}^{2}}}\\
        &= \frac{1}{2\ts}\pr{\sigma_{\xkr}^{2} - \sigma_{\xkpr}^{2} + \mu_{\xkr}^{2} - \mu_{\xkpr}^{2}} \ .
    \end{align}
\end{subequations}

\subsubsection{No-threshold feedback scheme for finite sampling frequency}

We vary the parameters $\alpha$ and $\psi$ to optimize the performance function~\eqref{eq:gen_ratchet_benefit} for input power~\eqref{eq:trap_astz_power_rel_coords} and rate of free-energy change~\eqref{eq:grav_astz_power_rel_coords}, averaged over distributions \eqref{eq:ss_natural_dxkt} and \eqref{eq:ss_natural_xkr}.
Equivalently, we choose the distributions parametrized by $\alpha$ and $\psi$ that maximize the general performance function,
\begin{align}
    \argmax_{\alpha,\psi}&\br{2\Gamma\epg_{\mrm{NT}}-2(1-\Gamma)\ept_{\mrm{NT}}}.
\end{align}
Here, the subscript ``NT'' denotes that these quantities are computed for the ``No-Threshold'' feedback scheme~\eqref{eq:natural_ansatz}. 
We obtain the optimal parameters
\begin{subequations}\label{eq:nt_optimal_params}
    \begin{align}
        \alpha^{*} &= 1\label{eq:nt_optimal_params1}\\
        \psi^{*} &= \frac{\Gamma+(1-\Gamma)\e^{-\ts}}{(1-\Gamma)(1+\e^{-\ts})}\dg\ .
    \end{align}
\end{subequations}
Substituting \eqref{eq:nt_optimal_params} into \eqref{eq:natural_ansatz}, the optimal no-threshold feedback rule for finite sampling frequency is 
\begin{align}
    \pr{\dlk^{*}}_{\mrm{NT}} = \xkpr + \frac{\Gamma+(1-\Gamma)\e^{-\ts}}{(1-\Gamma)(1+\e^{-\ts})}\dg\ ,\label{eq:full_non_cont_feedback}
\end{align}
producing optimal performance 
\begin{align}
    \mP_{\mrm{NT}}^{*} &=
    \frac{\pr{1-\e^{-2\ts}}}{\ts}\pr{1-\Gamma} + \frac{\tanh\tfrac{1}{2}\ts}{\ts}\frac{4\pr{\Gamma-\tfrac{1}{2}}^{2}}{1-\Gamma}\dg^{2}\ . 
    \label{eq:opt_fin_time_perf_func}
\end{align}

For continuous sampling (sampling period $\ts\to 0$, equivalently sampling frequency $\fs\to\infty$), the average rate of free-energy change~\eqref{eq:grav_astz_power_rel_coords} and trap power~\eqref{eq:trap_astz_power_rel_coords} simplify to
\begin{align}
    \epg_{\mrm{NT}}^{\infty} &= - \dg^{2} + \dg\pr{\frac{\psi}{\alpha}},\label{eq:short_time_pgrav}\\
    \ept_{\mrm{NT}}^{\infty} &= -1 - \dg\pr{\frac{\psi}{\alpha}} + \pr{\frac{\psi}{\alpha}}^{2}\ .\label{eq:short_time_ptrap}
\end{align}
The $\infty$ superscripts denote that these results are for infinite sampling frequency. 
For fixed $\dg$ and trade-off $\Gamma$, varying parameters $\alpha$ and $\psi$ to maximize~\eqref{eq:gen_ratchet_benefit},
\begin{align}
    \argmax_{\alpha,\psi}&\br{2\Gamma\epg_{\mrm{NT}}^{\infty}-2(1-\Gamma)\ept_{\mrm{NT}}^{\infty}}\ ,
\end{align}
gives arbitrary $\alpha$ and
\begin{align}
    \pr{\psi^{*}}^{\infty} &= \frac{\dg}{2\pr{1-\Gamma}}\alpha\ . \label{eq:nt_cont_fs_opt_psi}
\end{align}
Optimizing for the feedback scheme under the assumption of continuous sampling formally leaves undetermined the optimal $\alpha$; however, we choose $\alpha = 1$ to ensure consistency with the $\ts\to 0$ limit of the finite-sampling parameters~\eqref{eq:nt_optimal_params}. 
Thus, the optimal rule obtained from the continuous-sampling no-threshold feedback scheme~\eqref{eq:natural_ansatz} is
\begin{align}
    \pr{\dlk^{*}}_{\mrm{NT}}^{\infty} &= \alpha\sr{\xkpr + \frac{\dg}{2(1-\Gamma)}}\ , \label{eq:full_cont_sample_feedback_scheme} 
\end{align} 
producing 
\begin{align}
    \epg^{\infty}_{\mrm{NT}} &= \frac{\Gamma-\tfrac{1}{2}}{1-\Gamma} \dg^2 \label{eq:natural_pgrav}\\
    \ept^{\infty}_{\mrm{NT}} &=  \frac{\Gamma-\tfrac{1}{2}}{2\pr{1-\Gamma}^{2}}\dg^{2} - 1\ . \label{eq:natural_ptrap}
\end{align}
Figure~\ref{fig:greedy_natural_compare} displays these relations parametrically over the range $10^{-3} < \Gamma < 1$, for different values of $\dg$.
This continuous-sampling optimal rule~\eqref{eq:full_cont_sample_feedback_scheme} yields performance 
\begin{align}
    \pr{\mP_{\mrm{NT}}^{\infty}}^{*} = 2\sr{1-\Gamma + \frac{\pr{\Gamma-\tfrac{1}{2}}^{2}}{1-\Gamma}\dg^{2}}\ .
\end{align}

\subsubsection{Performance bounds under full extraction}
\label{subsubsec:performance_bounds_under_full_extraction}
We now show that performance for the no-threshold feedback scheme at finite sampling frequency is upper bounded by its value for continuous sampling, independent of the trade-off parameter $\Gamma$. 
In the optimal finite-sampling-frequency performance function~\eqref{eq:opt_fin_time_perf_func}, the factor $(1-\e^{-2\ts})/\ts$ in the first term monotonically decreases with the sampling period $\ts$ and goes to a maximum of unity in the continuous-sampling limit $\ts\to 0$. 
Similarly, $\tanh\pr{\ts/2}/\ts$ in the second term monotonically decreases with the sampling period, from a maximum of $1/2$ in the limit of $\ts\to 0$.
Put together, this yields
\begin{align}
    \mP_{\mrm{NT}}^{*} \le \pr{\mP_{\mrm{NT}}^{*}}^{\infty} \ .
\end{align}

\subsubsection{Verifying the optimality of the no-threshold scheme}
\label{subsubsec:hypothesis_verification}
To confirm that feedback schemes of the form~\eqref{eq:natural_ansatz} (i.e., with no threshold) are optimal, we now maximize the finite-sampling-frequency performance function~\eqref{eq:gen_ratchet_benefit} using the unconstrained feedback scheme~\eqref{eq:gen_scheme}, explicitly allowing a finite threshold $\xT$. 
A finite threshold presents analytical difficulties, as \eqref{eq:relative_pos_update} and \eqref{eq:relative_trap_update} produce non-Gaussian steady-state distributions with no tractable closed-form expressions.
We therefore numerically simulate many realizations of \eqref{eq:discrete_od_langevin} for each set of feedback parameters $\{\alpha,\psi,\xT\}$ and the unconstrained feedback scheme~\eqref{eq:gen_scheme}, to estimate the rate of free-energy change~\eqref{eq:nat_pgrav_def} and trap power~\eqref{eq:nat_ptrap_def}.
We then use stochastic gradient ascent~\cite{Wales1997} in the space of feedback parameters to find the optimal set that maximizes performance~\eqref{eq:gen_ratchet_benefit} for fixed trade-off parameter $\Gamma$.
(Appendix~\ref{sec:grad_ascent} provides numerical details.)
As we have already seen that performance is generally maximized for high sampling frequencies, we use $\fs = 1000$ in the simulations.

\begin{figure}[htbp]
    \centering
    \includegraphics[clip, width=\linewidth]{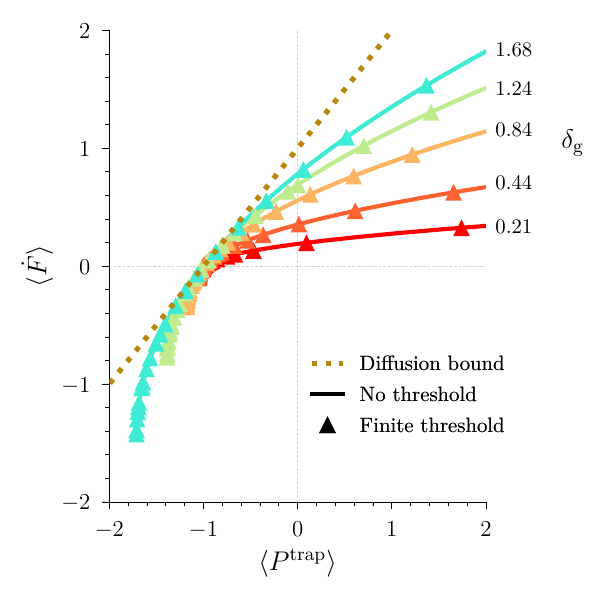}
    \caption{
        Pareto performance frontiers for continuous-sampling unconstrained feedback schemes with no threshold 
        (\ref{eq:natural_pgrav}, \ref{eq:natural_ptrap}) (solid curves) or permitting finite threshold and optimized using numerical gradient ascent (triangles).
        Gold dotted line: diffusion bound on the net output power of $\epn = 1$.
        }
    \label{fig:greedy_natural_compare}
\end{figure}

The squares in Fig.~\ref{fig:greedy_natural_compare} show $\epg$ and $\ept$ for the rules obtained by gradient ascent.  
These values generally agree well with the frontiers of the no-threshold feedback scheme.
This finding is consistent with the conclusion that the dominant strategy, for any trade-off between $\epg$ and $\ept$, is one that uses all of the available fluctuations and hence has no threshold.

\subsubsection{Feedback cooling}

Under the unconstrained feedback scheme, for each $\dg$ the Pareto frontier at trade-off parameter $\Gamma=1/2$ (where performance equals net output power) passes through $(\epg = 0, \ept = -1)$.
For this case, the optimal finite-sampling-frequency feedback rule~\eqref{eq:full_non_cont_feedback} dictates that $\psi = \dg$, implying
\begin{subequations}\label{eq:max_pnet_opt_naive_rule}
    \begin{align}
        \dlk &= \xkpr + \dg\\
        \lkp - \dg &= \xkp\ .
    \end{align}
\end{subequations}
In the second line, we used the definitions for $\dlk$~\eqref{eq:trap_update} and $\xkpr$~\eqref{eq:relative_pos_update}.
This feedback rule moves the minimum of the \emph{total} potential (located at $\lkp-\dg$) to the last measured particle position $\xkp$, thereby extracting at each measurement the particle's total potential energy.

This feedback rule gives $\mu_{\xkpr} = \mu_{\xkr}$, meaning zero steady-state average rate of free-energy change ($\epg_{\mrm{NT}}~=~0$).
Since \eqref{eq:max_pnet_opt_naive_rule} forces the minimum of the total potential to follow the particle at each time step, the particle is as likely to fluctuate up or down in the subsequent interval between measurements.
The minimum of the total potential therefore undergoes similar unbiased Brownian motion, on average producing no free-energy output.

This feedback scheme only extracts work via the trap and mimics an overdamped version of \emph{feedback cooling}~\cite{Kim2007,Horowitz2014}, in which the variance of the particle distribution is reduced relative to the equilibrium distribution. 
(In fact, \eqref{eq:max_pnet_opt_naive_rule} imposes that $\ev{\pr{\xkr}^{2}} = 0$.)
Additionally, $\ev{\pr{\xkpr + \dg}^{2}} = \sigma_{\ts}^{2} = 1-\e^{-2\ts}$, which implies steady-state average power 
\begin{align}
    \ept_{\mrm{NT}} = -\frac{1-\e^{-2\ts}}{2\ts}\ . \label{eq:m2_fdt_expr}
\end{align}
The net output power in the limit of continuous sampling is then upper bounded by
\begin{align}
    \epn_{\mrm{NT}}^{\infty} = -\ept_{\mrm{NT}}^{\infty} = 1\ . \label{eq:pnet_bound} 
\end{align}
In dimensional units, the rule that extracts all available energy at each measurement produces finite net output power $\epn_{\mrm{NT}} = 1\, \kT/\trelo$, even with continuous measurements.
This limit is independent of $\dg$; as a result, all Pareto frontiers (across different $\dg$'s) in Fig.~\ref{fig:greedy_natural_compare} cross at the same point where $\epg_{\mrm{NT}}^{\infty}=0$ and $\ept_{\mrm{NT}}^{\infty} = -1$.

\subsection{Practical-storage feedback schemes}
\label{subsec:practical_storage_schemes}

We now restrict ourselves to the case where negative trap work cannot be stored.
We therefore consider feedback schemes, either with the constraint that no trap (input) work be done~\eqref{eq:pnas_rule} (zero-work feedback scheme) or with the 
modified rule set~\eqref{eq:strict_scheme} (nonnegative-work feedback scheme) allowing nonnegative trap work.

\subsubsection{Zero-work feedback scheme}
\label{subsubsec:maxwell_demon_feedback_scheme}

In \cite{Saha2021}, we optimized net output power when the engine is constrained to operate as a ``Maxwell demon'' that does no trap work.
The constraint $\wt_k = 0$ was imposed for each ratchet event, thus requiring \eqref{eq:pnas_rule}.
This rule enforces a symmetry on the process that makes the self-consistent equation for the steady-state distributions $\pi\pr{\xkpr}$ and $\tilde{\pi}\pr{\xkr}$ numerically solvable~\cite[SI App. E]{Saha2021}. 

\begin{figure}[ht!]
    \centering
    \includegraphics[clip, width=\linewidth]{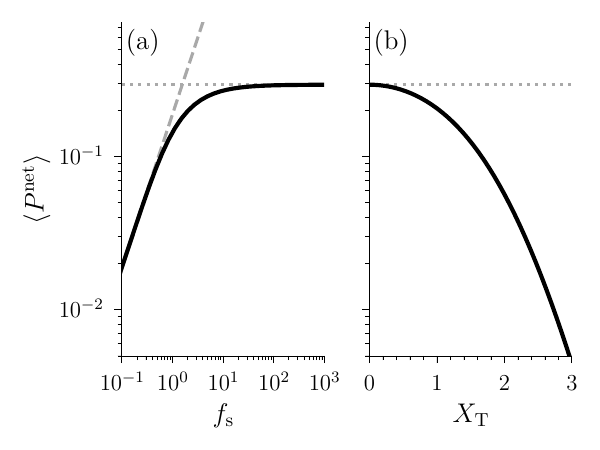}
    \caption{
        Net output power $\epn$ as a function of sampling frequency and threshold for zero-work feedback scheme.
        (a) $\epn$ as a function of sampling frequency $\fs$ for $\dg=0.84$ and $\xT = 0$. 
        Dashed gray line: near-equilibrium prediction~\eqref{eq:quasistatic_MD_pnet}.
        (b) $\epn$ as a function of threshold $\xT$ for $\dg=0.84$ and $\fs = 1000$.
        Dotted horizontal gray line: predicted continuous-sampling power~\eqref{eq:pow_theory}.
        }
    \label{fig:pnas_fig}
\end{figure}

The net output power for this scheme is simply the rate of free-energy change and can be computed from \eqref{eq:grav_astz_power_rel_coords}, where the means $\mu_{\xkpr}$ and $\mu_{\xkr}$ are computed by averaging over the steady-state joint distribution $\pi\pr{\xkpr,\xkr}$ of the relative coordinates. 
Figure~\ref{fig:pnas_fig}a shows that the output power saturates at large sampling frequency. 
That net output power is maximized at high sampling frequencies is consistent with the more general finding of Sec.~\ref{subsubsec:performance_bounds_under_full_extraction} that performance is maximized for continuous sampling.

In the limit of low sampling frequency, the system equilibrates between each measurement/feedback step, reaching the Gibbs-Boltzmann position distribution~\eqref{eq:eq_dist}.
In this limit, the average work extracted is determined from the feedback rule~\eqref{eq:pnas_rule},
\begin{align}
    \ev{P^{\mrm{net}}_{\fs\to 0}}  = \fs \br{\sqrt{\frac{2}{\pi}}\dg\e^{-\dg^{2}/2} + \dg^{2}\sr{\erf\pr{\frac{\dg}{\sqrt{2}}}-1}}.\label{eq:quasistatic_MD_pnet}
\end{align}
The power is linear in the sampling frequency, as shown in Fig.~\ref{fig:pnas_fig}a.

The maximal net power in Fig.~\ref{fig:pnas_fig}a, is achieved in the limit of large sampling frequency (vanishing sampling time).
(Although experimental sampling times are always finite, these times can be much faster than system time scales.)
Sending sampling time to zero eases calculation of the diffusive particle's mean first-passage time (MFPT), the time to first cross the threshold $\xT$ when starting at $-\xR$~\cite{Hanggi1990,Pontryagin1933}, 
\begin{align}
    \tmfp = \int_{-\xR}^{\xT}\dd{x'}\int_{-\infty}^{x'}\dd{x''} \exp\sr{V(x')-V(x'')}\label{eq:MFPT} \ ,
\end{align}
for total potential $V(x) = \frac{1}{2}x^{2}+\dg x$.
The feedback rule~\eqref{eq:pnas_rule} sets the threshold to the start position, $\xR=\xT$, giving net output power
\begin{align}
    \epn^{\infty} = \frac{2\xT}{\tmfp} \dg \ .
\end{align}
Figure~\ref{fig:pnas_fig}b shows that empirically the net output power is maximized as $\xT \to 0$.
Thus, the \emph{optimal zero-work feedback rule} is
\begin{align}
    \dlk = \Theta\pr{\xkpr}2\xkpr\ .\label{eq:opt_MD_rule}
\end{align}
Inspired by the empirically observed maximal net output power and vanishing threshold, we make a small-$\xT$ expansion in \eqref{eq:MFPT}, giving
\begin{align}
    \tmfp_{\xT\to0}^{\infty} \approx \sqrt{2\pi}\e^{\dg^{2}/2}\sr{1+\erf\pr{\frac{\dg}{\sqrt{2}}}}\xT\label{eq:MFPT_opt}
\end{align}
to first-order in the threshold, yielding net output power
\begin{align}
    \epn_{\xT\to 0}^{\infty} &= \sqrt{\frac{2}{\pi}}\dg\e^{-\dg^{2}/2}\sr{1+\erf\pr{\frac{\dg}{\sqrt{2}}}}^{-1} \ . \label{eq:pow_theory}
\end{align}
This result for the net output power reprises the derivation in \cite{Saha2021} and has also been found previously~\cite{Park2016} via a different method.

\begin{figure}[htbp]
    \centering
    \includegraphics[clip, width=\linewidth]{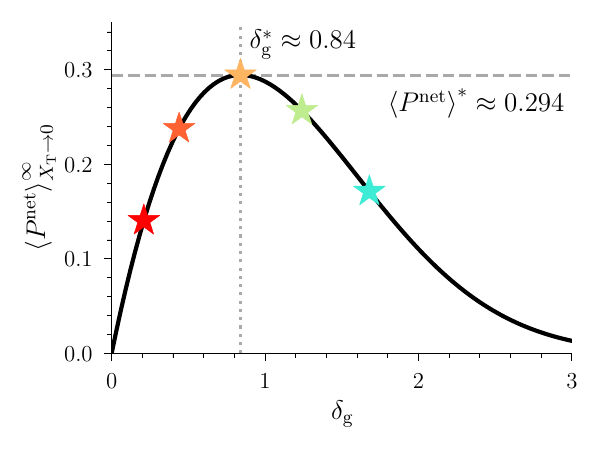}
    \caption{
        Net output power $\epn$ as a function of scaled effective mass $\dg$ for the optimal zero-work feedback rule~\eqref{eq:opt_MD_rule}.
        Solid curve: theoretical approximation in the $\Gamma\to 0$ limit~\eqref{eq:pow_theory}.
        Stars: $\dg$'s shown in Fig.~\ref{fig:strict_max}.
        Dotted vertical gray line: $\dg\approx 0.84$ that maximizes $\epn$.
        Horizontal dashed gray line: maximal $\epn^* \approx 0.294$.
        }
    \label{fig:pnas_fig2}
\end{figure}

Having optimized the net output power with respect to the sampling frequency and the threshold, we now optimize with respect to the scaled effective mass $\dg$. Figure~\ref{fig:pnas_fig2} shows a maximum net extraction power $\epn^{*} \approx 0.294$ at intermediate $\dg^{*}\approx 0.84$.
This maximum arises from the competition between increased scaled effective mass increasing potential energy yet also increasing the time required to fluctuate beyond the threshold.

\subsubsection{Nonnegative-work feedback scheme} 
\label{subsubsec:strict_feedback_scheme}
\begin{figure}[htbp]
    \centering
    \includegraphics[clip, width=\linewidth]{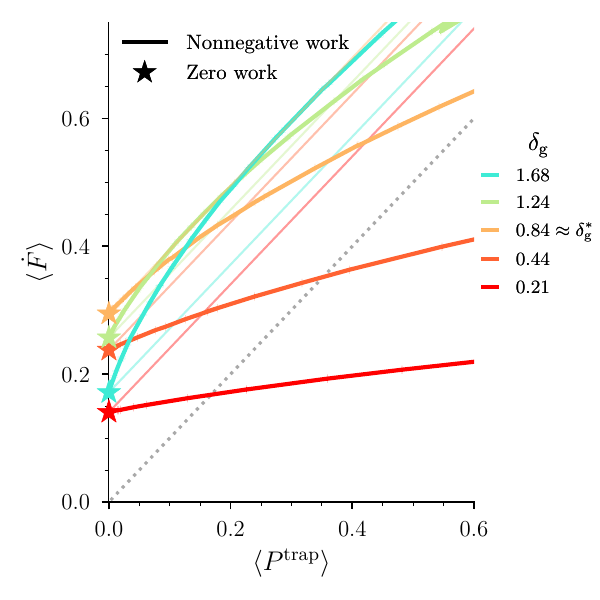}
    \caption{
        Pareto frontiers (solid curves) under the nonnegative-work (practical-storage) feedback scheme, for different scaled effective masses $\dg$ (different colors). 
        Stars: optimal zero-work feedback rule (Fig.~\ref{fig:pnas_fig2}). 
        Thin solid lines: net output power $\epn$ equal to corresponding star. 
        Dotted gray line: $\epn=0$. 
        }
    \label{fig:strict_max}
\end{figure}

As with the unconstrained feedback scheme, the nonlinearities in the nonnegative-work feedback scheme~\eqref{eq:strict_scheme} make analysis difficult; consequently, we numerically estimate Pareto frontiers using stochastic gradient ascent (App.~\ref{sec:grad_ascent}) on the space of parameters $\{\alpha,\psi,\xT\}$.
Guided by the analysis in Sec.~\ref{subsubsec:performance_bounds_under_full_extraction}, which revealed an advantage to frequent sampling, and similar to Sec.~\ref{subsubsec:hypothesis_verification}, we use sampling frequency $\fs = 1000$. (Appendix~\ref{sec:sampling_freq_variation} explores the effect of varying sampling frequency.)
Figure~\ref{fig:strict_max} shows the Pareto frontiers found by gradient ascent.

In the limit of small trade-off parameter ($\Gamma\to 0$), the performance~\eqref{eq:gen_ratchet_benefit} becomes independent of the average rate of free-energy change $\epg$ and proportional only to the negative of the trap power, $\mP\propto -\ept$: maximizing performance is equivalent to minimizing the trap input power.
The practical-storage schemes impose that $\ept \ge 0$ and thus include as a special case the zero-work feedback scheme from Sec.~\ref{subsubsec:maxwell_demon_feedback_scheme}, which imposes $\ept = 0$.
There is therefore a correspondence (denoted by the colored stars in Figs.~\ref{fig:pnas_fig2} and \ref{fig:strict_max}) between the optimized zero-work scheme and the optimized nonnegative-work feedback scheme for $\Gamma\to 0$. 
($\Gamma=0$ presents complications: for a given $\dg$, all $\xT > 0$ admit optimal solutions that achieve $\ept=0$, with $\epg$ varying between 0 and the optimal value at finite-but-small $\Gamma$.) 

For $\dg < \dg^{*} \approx 0.84$, Fig.~\ref{fig:strict_max} shows that other rules, generally with nonzero trap work, do not increase the net output power beyond that for $\Gamma\to 0$. 
In contrast, for $\dg > \dg^{*}$, net output power can be increased with an input of trap power, with the Pareto frontiers upper bounded by the net output power associated with the zero-work feedback scheme~\eqref{eq:pow_theory} evaluated for the optimal mass $\dg^{*}$.
That is, for heavier particles with effective mass $m$ leading to $\dg > \dg^{*}$, investing work as input is more than compensated by increasing the free-energy change extracted as output.

To understand the physical mechanism underlying optimal feedback rules that maximize the net output power (i.e., for $\Gamma = 1/2$), we map the process imposed by the nonnegative-work feedback scheme to the optimal zero-work feedback rule: we choose the parameters $\{\alpha,\psi,\xT\}$ such that the nonnegative-work feedback scheme is equivalent to that of the optimal zero-work rule.
To do this, we rewrite the nonnegative-work feedback scheme~\eqref{eq:strict_scheme} and optimal zero-work feedback rule~\eqref{eq:opt_MD_rule} in the frame that is co-moving with the total potential. 
By comparing the nonnegative-work feedback scheme to the optimal zero-work feedback rule, we identify the parameters for the nonnegative-work feedback scheme that makes the resulting feedback rules the same (see App.~\ref{sec:mapping} for details), yielding
\begin{subequations}\label{eq:opt_net_output_params}
    \begin{align}
        \alpha^{*} &= 2\\
        \psi^{*} &= 2\pr{\dg-\dg^{*}} \label{eq:opt_net_output_params2}\\
        \xT^{*} &= -\frac{1}{2}\psi^{*} \label{eq:opt_net_output_params3}\ .
    \end{align}
\end{subequations}
This choice of parameters also reproduces the maximum value of the net output power~\eqref{eq:pow_theory}, 
\begin{align}
    \epn^{*} \equiv \sqrt{\frac{2}{\pi}}\dg^{*}\e^{-\pr{\dg^{*}}^{2}/2}\sr{1+\erf\pr{\frac{\dg^{*}}{\sqrt{2}}}}^{-1}\ ,\label{eq:pow_theory_max}
\end{align}
for all $\dg > \dg^{*}$ (Appendix~\ref{sec:mapping}).

\begin{figure}[htbp]
    \centering
    \includegraphics[clip, width=\linewidth]{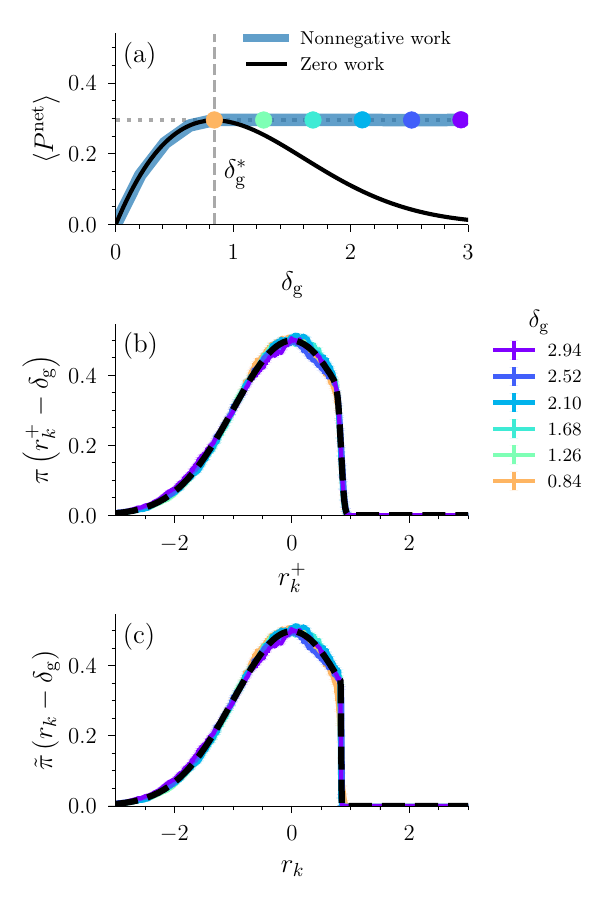}
    \caption{
        Mapping the nonnegative-work scheme to the optimal zero-work feedback rule reproduces the performance bound for $\dg \gtrsim \dg^{*}$.
        (a) Optimal net output power $\epn$ as a function of scaled effective mass $\dg$. 
        Blue wide curve: nonnegative-work feedback. 
        Black thin curve: zero-work feedback from Fig.~\ref{fig:pnas_fig2}.
        Circles: net output power associated with the steady-state distributions in (b, c).
        Gray dashed vertical line: scaled effective mass $\dg^{*}$ that maximizes the output power in the optimal zero-work rule.
        (b, c) Steady-state distributions of shifted relative coordinates $\xkpr + \dg$ and $\xkr + \dg$, respectively. 
        Dashed black curve: shifted steady-state distribution of the optimal zero-work feedback scheme for $\dg=\dg^{*}$.
        }
    \label{fig:opt_strict_scheme_dg_var}
\end{figure}

To confirm this analysis, we fix $\alpha = 2$, and, for a variety of $\dg$, perform stochastic gradient ascent to find $\psi$ and $\xT$ that (locally) optimize net power (i.e., performance with $\Gamma = 1/2$).
Figure~\ref{fig:opt_strict_scheme_dg_var}a shows that the resulting net power reproduces the output of the optimal zero-work feedback scheme~\eqref{eq:pow_theory} when $\dg < \dg^{*}$ [Fig.~\ref{fig:pnas_fig2}], further supporting that input trap power cannot compensate for a small effective mass. 
However, when $\dg \gtrsim \dg^{*}$, the numerically optimized net power uniformly achieves the maximal (over all $\dg$) net output power 
$\epn^{*}\approx 0.294$~\eqref{eq:pow_theory_max}. 
Figures~\ref{fig:opt_strict_scheme_dg_var}b and c verify that the optimal nonnegative-work feedback rules produce steady-state distributions 
of the (shifted) relative coordinates that match those of the optimal zero-work feedback rule.

We show (App.~\ref{sec:work_distributions}) that this empirical equality between steady-state distributions, shown in Figs.~\ref{fig:opt_strict_scheme_dg_var}b and c, implies that the net output power distributions of the optimal nonnegative-work feedback rule and the optimal zero-work feedback rule are also equal. Thus, this mapping recovers not only the mean net output power but the full distribution of the net output power as well.

\begin{figure}
    \centering
    \includegraphics[clip, width=\linewidth]{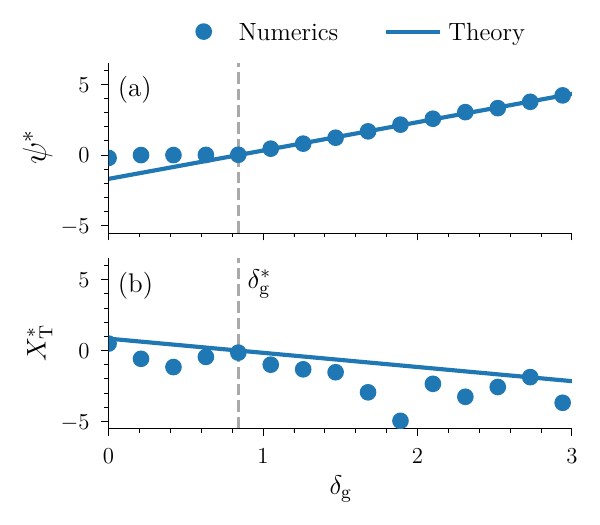}
    \caption{
        Nonnegative-work-scheme parameters associated with maximal net output power. 
        Optimal offset $\psi^{*}$ (a) and threshold $\xT^{*}$ (b) for nonnegative-work feedback scheme with $\Gamma = 1/2$ and fixed proportionality $\alpha^{*} = 2$, as function of scaled effective mass $\dg$.  
        Blue points: optimal parameters found numerically by gradient ascent.
        Solid blue lines: theory~\eqref{eq:opt_net_output_params} from mapping the nonnegative-work feedback scheme to the optimal zero-work (ZW) feedback rule.
        Dashed gray vertical line: scaled effective mass $\dg^{*}\approx 0.84$ optimizing ZW performance.
        }
    \label{fig:opt_net_output_params_strict}
\end{figure}

Furthermore, when its rules are optimized for net output power, if the nonnegative-work feedback scheme does realize the same process as the optimal zero-work feedback scheme, then \eqref{eq:opt_net_output_params} should allow characterization of the optimal parameters resulting from this restricted (fixed-$\alpha$) stochastic gradient ascent. 
Indeed, the empirical optimal values of $\psi^{*}$ in Fig.~\ref{fig:opt_net_output_params_strict}a are indistinguishable from the theoretical expectation~\eqref{eq:opt_net_output_params2} for $\dg > \dg^{*}$.
Figure~\ref{fig:opt_net_output_params_strict}b shows that \eqref{eq:opt_net_output_params3} provides an upper bound for the possible thresholds $\xT$, which is expected due to the extra constraints of the nonnegative-work feedback scheme~\eqref{eq:strict_scheme}.
Specifically, the Heaviside function in the proposal~\eqref{eq:strict_scheme1} imposes $\xT \le \xkpr$, while the Heaviside function imposing non-negative trap work in \eqref{eq:strict_scheme2} requires $-|\xkr| \le \xkpr \le |\xkr|$; hence, nonnegative-work feedback rules with $\xT<-|\xkr|$ have the same performance as those with $\xT = -|\xkr|$.

\begin{figure}
    \centering
    \includegraphics[clip, width=\linewidth]{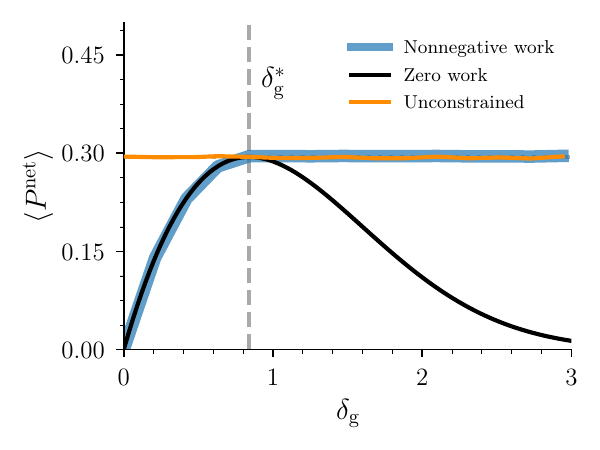}
    \caption{ 
    Net output power of the unconstrained and nonnegative-work feedback schemes, both using~\eqref{eq:opt_net_output_params}, as well as the optimal zero-work feedback rule. 
    Black thin curve and blue wide curve: same as in Fig.~\ref{fig:opt_strict_scheme_dg_var}a. 
    Orange curve: unconstrained feedback rule obtained by mapping the nonnegative-work feedback scheme to the optimal zero-work feedback rule~\eqref{eq:opt_net_output_params}.
    Dashed gray vertical line: same as in Fig.~\ref{fig:opt_strict_scheme_dg_var}a.
    }
    \label{fig:gravitation_limits}
\end{figure}

Figures~\ref{fig:strict_max} and \ref{fig:opt_strict_scheme_dg_var} show a sharp transition in optimal strategies at $\dg = \dg^{*}$, from strategies that do not input work when $\dg < \dg^{*}$ to strategies that input work for $\dg > \dg^{*}$. 
To understand this transition, we apply to the unconstrained feedback scheme~\eqref{eq:gen_scheme} the rules obtained from mapping the nonnegative-work scheme to the optimal zero-work feedback rule~\eqref{eq:opt_net_output_params} and compute the corresponding net output powers.
Figure~\ref{fig:gravitation_limits} shows that the unconstrained scheme can uniformly recover the maximum $\epn^{*}$ of the optimal zero-work feedback scheme for all values of $\dg$. 
Conversely, as seen previously in Fig.~\ref{fig:opt_strict_scheme_dg_var}, the rules~\eqref{eq:opt_net_output_params} applied to the nonnegative-work feedback scheme recover the maximum net output power only for $\dg > \dg^{*}$: for $\dg < \dg^{*}$, there are no nonnegative-work-scheme strategies that increase the net output work without negative trap work. 
Thus, the restrictions of the nonnegative-work feedback scheme prevent it from realizing the maximum net output power.

\begin{figure} 
    \centering 
    \includegraphics[clip, width=\linewidth]{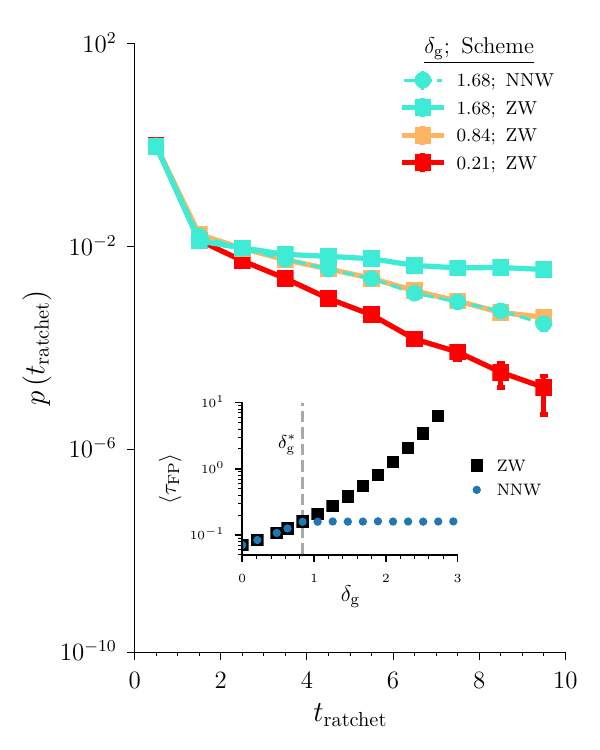}
    \caption{
    Ratchet-time distributions, each computed for a trajectory with protocol duration $t_{\mrm{prot}} = 5000$ and sampling frequency $\fs=1000$.
    Different colors denote different $\dg$. 
    Squares and solid connecting curves: distributions for the optimal zero-work (ZW) feedback rule~\eqref{eq:opt_MD_rule}. 
    Points and dashed connecting curve: distribution for the optimal nonnegative-work (NNW) feedback rule~\eqref{eq:opt_net_output_params}. 
    Error bars denote standard error of the mean. 
    The ratchet-time distribution for optimal ZW feedback rule with $\dg = 0.84$ is indistinguishable from the ratchet-time distribution for the NNW feedback scheme using the rules~\eqref{eq:opt_net_output_params} with $\dg = 1.68$.
    Inset: mean first-passage time (computed from trajectories) as a function of scaled effective mass $\dg$, for optimal ZW feedback scheme (squares) and for NNW feedback scheme (points). 
    Dashed vertical line denotes $\dg^{*} \approx 0.84$ optimizing ZW performance.
    }
    \label{fig:ratchet_time_distr}
\end{figure}

Supplying input work shifts the total potential such that the particle experiences stronger upward forces, which should reduce the time to diffuse past the threshold. 
To confirm this, we measure the ratchet-time distribution (i.e., the time between ratchet events $t_{\mathrm{ratchet}}$) during a long simulated trajectory of \eqref{eq:discrete_od_langevin}
\footnote{We construct the ratchet-time distribution by computing, for a given trajectory, the time $t_{\mathrm{ratchet}}$ between trap-center movements along that trajectory. We histogram these times (properly normalized) to obtain the densities in Fig.~\ref{fig:ratchet_time_distr}.}.
We collect measurements at a sampling frequency $\fs=1000$, approximating the continuous-sampling limit.
As such, we can view the ratchet-time distributions as approximately equivalent to the first-passage time distributions since, at this fast-sampling limit, we accurately detect when the particle passes the threshold $\xT$ for the first time.

Figure~\ref{fig:ratchet_time_distr} shows the ratchet-time distributions for the optimal zero-work feedback rule~\eqref{eq:opt_MD_rule} that does not input work and for the optimal nonnegative-work feedback rule~\eqref{eq:opt_net_output_params} that does input work.
For the optimal zero-work feedback rule, the proportion of events associated with very long first-passage times increases with scaled effective mass $\dg$. 
However, for a given $\dg$, the proportion of these long excursions away from the threshold can be reduced by an input of work. 
This causes the first-passage time distribution for higher $\dg$ to match that of the optimal $\dg^{*}$, thus equalizing the mean first-passage times as seen in the inset of Fig.~\ref{fig:ratchet_time_distr}.
Thus, the input of work serves to reduce the mean first-passage time for scaled effective masses $\dg > \dg^{*}$, thereby enhancing the net output power of the optimal nonnegative-work feedback rules relative to the optimal zero-work feedback rule [Fig.~\ref{fig:opt_strict_scheme_dg_var}a; Fig.~\ref{fig:gravitation_limits}].

\section{Conclusion} 
\label{sec:conclusion}

We have examined the performance limits of a simple realization of an information engine comprised of a massive colloidal particle immersed in an isothermal reservoir and diffusing in a harmonic potential under gravity. 
This information engine can perform work against gravity and thus store free energy that can be harnessed later.
Guided by practical considerations, we studied this engine's performance under different feedback schemes, each with a different constraint on the allowable energy extraction. 
Table~\ref{tbl:summary_table} provides a summary of our results.

Our analysis of the unconstrained scheme, where the information engine can store all of the potential energy (both as work and as free-energy change) that is input into the particle by thermal fluctuations, shows that the net rate of energy extraction (net output power) is limited to $1\ \kT/\trelo$ across all values of scaled effective mass $\dg$ (Fig.~\ref{fig:greedy_natural_compare}). 
Closer analysis of the unconstrained rule~\eqref{eq:max_pnet_opt_naive_rule} that achieves this bound reveals that energy extraction happens entirely via the trap potential~\eqref{eq:pnet_bound}. 
This limit is independent of the sampling frequency; thus, continuously collecting information provides diminishing returns.
Furthermore, we find that the optimal unconstrained rules~\eqref{eq:natural_ansatz} do not use a threshold to selectively filter for a subset of fluctuations.

Conversely, in cases where the information engine cannot store energy as work via the trap (and thus does not benefit from negative trap work), the net output power is uniformly bounded~\eqref{eq:pow_theory_max} by $\approx 0.294\ \kT/\trelo$ in physical units (Fig.~\ref{fig:gravitation_limits}).
This bound is more restrictive than the bound on the unconstrained schemes. 
Thus, storing energy as work via a nonlinear potential (the quadratic trap) is more effective than storing energy as a free-energy change via a linear potential (the gravitational potential).

In practical-storage schemes, there is a sharp transition in optimal strategies (Fig.~\ref{fig:strict_max}) between optimal strategies for $\dg < \dg^{*}$ that do not input work but cannot reach the maximal $\epn^{*} = 0.294\ \kT/\trelo$, and optimal strategies for $\dg > \dg^{*}$ that do input work and realize the bound on the maximum net output power. 
This sharp change in optimal strategy as a function of system parameters is a common phenomenon in control theory~\cite{bechhoefer_book2021,Solon2018}.

\begin{table*}[]
    \normalsize
    \setlength{\tabcolsep}{0.5em}
    \setlength{\extrarowheight}{1.0em}
    \renewcommand{\arraystretch}{0.7}
    \begin{tabular}[c]{|>{\centering\arraybackslash}m{0.16\textwidth}|c|c|p{0.1\textwidth}<{\centering}|p{0.15\textwidth}|p{0.19\textwidth}|}
        \hline
        Scheme & $\wt_k$ & $\displaystyle \argmax_{\dg}\epn$ &
        \begin{tabular}[c]{@{}c@{}}
            $\epn^{*}$\\
            $(\kT/\trelo)$
        \end{tabular} &
        \multicolumn{2}{c|}{
        \begin{tabular}[c]{@{}c@{}}
                Behavior under optimal rules
        \end{tabular}} \\ [1.5em] \hline
        \vspace*{0.4cm} Unconstrained &
        any &
        $\forall\ \dg$ & \centering
        $1$ &
        \multicolumn{2}{l|}{
            \begin{tabular}[c]{@{}l@{}}
                \textbullet~captures all fluctuations\\ 
                \textbullet~all output via trap
            \end{tabular}
        } \\[2.1em] \hline
        \vspace{0.9cm} Zero work &
        $= 0$ &
        $\approx 0.84$ &
        \vspace*{0.3cm}
        \multirow{2}{*}{
                $\approx 0.294$
            } & 
        \vspace*{0.0125cm}
        \multirow{2}{0.14\textwidth}{rectifies 
        all ``up'' fluctuations and only ``up'' fluctuations
        } &
        \vspace*{-1cm} greater mass trades off greater work with longer $\tmfp$ 
        \\ [0.2em] \cline{1-3} \cline{6-6} 
        \vspace{0.5cm} Nonnegative work &
        $\ge 0$ &
        $\gtrsim 0.84$ &
        &
        &
        \vspace*{-0.8cm} input work to reduce $\tmfp$ \\[1.0em]
        \hline
    \end{tabular}
    \caption{Summary of results for the unconstrained~\eqref{eq:gen_scheme}, the zero-work~\eqref{eq:pnas_rule}, and the nonnegative-work~\eqref{eq:strict_scheme} feedback schemes (rows): permissible trap-work increments $\wt_k$; $\dg$ range that optimizes $\epn$; maximal net output power $\epn^*$; and qualitative behavior under optimal rules (columns).
    }
    \label{tbl:summary_table} 
\end{table*}

We find that the input of work, for $\dg > \dg^{*}$, is used to reduce the proportion of trajectories that wander away from the threshold for a long time, thus speeding up the overall engine operation (Fig.~\ref{fig:ratchet_time_distr}).
We also find that the inability for the practical-storage schemes to store energy from the trap severely limits engine output when $\dg < \dg^{*}$, as the strategies available to increase net output power in this regime require storing energy via the trap (Fig.~\ref{fig:gravitation_limits}).

The limit on output power for the practical-storage feedback schemes arises because the engine can only rectify ``up'' fluctuations.
Information collected about the ``down'' fluctuations is wasted, since there is no feedback response.
This contrasts with the unconstrained scheme, where the engine acts on both ``up'' and ``down'' fluctuations, thus exploiting all the information it gathers and producing a much higher net power that is limited only by diffusion (entering through the relaxation time $\tau_\textrm{R}$).

Our observation that input energy can increase information-engine performance has been seen in other systems. 
Schmitt et al.~\cite{schmitt2015} study an information engine similar to ours, with an overdamped particle diffusing freely in a piecewise-linear potential coupled to a feedback control system that rectifies fluctuations by moving the potential a fixed distance $L$ if the particle is measured beyond some pre-defined threshold $\xT$. In the parameter space they explored, they find---as we do---that performance is maximized when the engine both uses information and inputs work.

We primarily focused on maximizing engine performance, as quantified by the trade-off between rate of free-energy change and trap input power.
We have not specified any details of the apparatus processing the measurements (the measurement device) and evaluating the feedback rule (the feedback controller); however, we assume that continuous sampling (i.e., the ability to sample much faster than characteristic dynamical time scales) and noise-free measurements are possible.
In practice, the latter amounts to requiring that the uncertainty of position measurements be significantly less than the trap length scale $\sigma$.
While there are a variety of ways of estimating the thermodynamic costs of information processing~\cite{Horowitz2014}, all lead to vanishing efficiency when applied to our setup.
It would be interesting to explore what optimal feedback rules arise when the information cost is included in the performance function.

Finally, although we were inspired by the recent experiment in \cite{Saha2021}, our results should hold more generally for systems under the influence of a joint quadratic-plus-linear potential. 
Recent experiments~\cite{Chang2021}, for example, have realized electronic circuits that can be configured such that the charge dynamics in the circuit are governed by \eqref{eq:od_langevin}. 
With such a setup, one could implement the feedback schemes we have outlined here and test our predicted performance limits. 

\begin{acknowledgments}
The authors thank Tushar Saha, Steven Blaber (SFU Physics), and Dorian Daimer (UH Manoa Physics) for helpful discussions and comments on the manuscript.
This research was supported by grant FQXi-IAF19-02 from the Foundational Questions Institute Fund, a donor-advised fund of the Silicon Valley Community Foundation. 
Additional support was from Natural Sciences and Engineering Research Council of Canada (NSERC) Discovery Grants (D.A.S.\ and J.B.), a Tier-II Canada Research Chair (D.A.S.), an NSERC Undergraduate Summer Research Award, a BC Graduate Scholarship, and an NSERC Canadian Graduate Scholarship -- Masters (J.N.E.L.).
Computational support was provided by WestGrid and Compute Canada Calcul Canada.
\end{acknowledgments}

\appendix

\section{Dissipative feedback scheme}
\label{sec:diss_feedback}

\begin{figure}
    \centering 
    \includegraphics[trim = 1 1 1 1, clip, width=\linewidth]{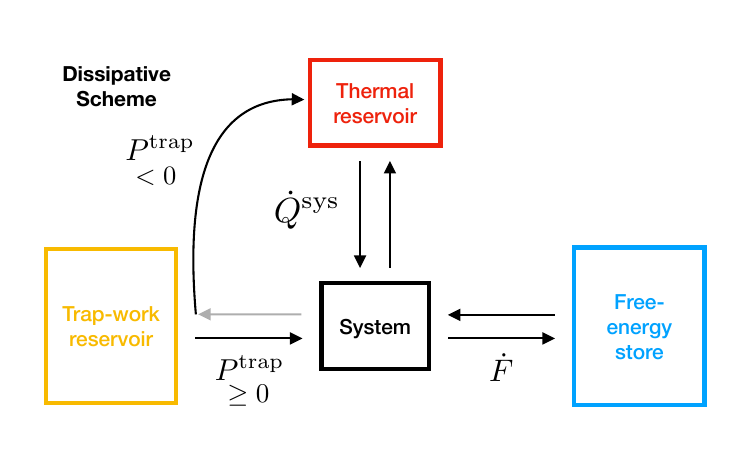}
    \caption{
        Schematic of the energy flows
        associated with the dissipative feedback scheme,
        among the system, the free-energy store, the trap-work reservoir, and the thermal reservoir.
        }
    \label{fig:diss_thermo_diagram}
\end{figure}

Another way to address the practical considerations, in addition to the zero-work (Sec.~\ref{subsubsec:maxwell_demon_feedback_scheme}) and nonnegative-work (Sec.~\ref{subsubsec:strict_feedback_scheme}) feedback schemes discussed in the main text, is to modify the accounting of the trap input work for the case where the energy which flows from the system to the trap-work reservoir is not storable and thus is ultimately dissipated into the thermal reservoir (Fig.~\ref{fig:diss_thermo_diagram}), providing no direct performance improvement to the engine. 
To reflect this, we modify the definition of the trap input power to 
\begin{align}
    \eptm &= \frac{1}{t_{\mrm{prot}}}\sum_{k}\Theta\sr{\wt_{k}}\wt_{k}.\label{eq:ptrap_blind}
\end{align}
This differs from \eqref{eq:nat_ptrap_def} in that the Heaviside function inside the summation ensures that only positive contributions are counted towards the total trap input power.
This \emph{dissipative} scheme therefore dissipates the energy associated with moves that have negative trap work.

\begin{figure}[!htbp]
    \centering
    \includegraphics[clip, width=\linewidth]{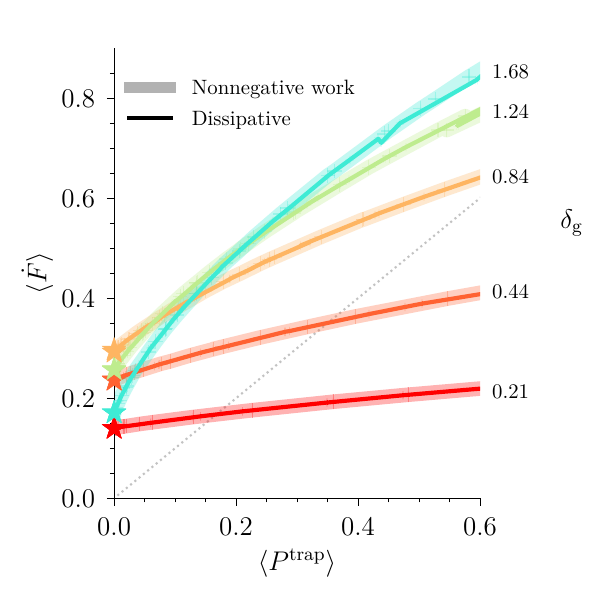}
    \caption{ 
        Comparison of Pareto frontiers for nonnegative-work and dissipative feedback schemes, for different scaled effective masses $\dg$ (different colors). 
        Light wide curves: Pareto frontiers for the nonnegative-work feedback scheme, same as in Fig.~\ref{fig:strict_max}.
        Dark narrow curves: Pareto frontiers for the dissipative feedback scheme.
        The Pareto frontiers for the two feedback schemes are indistinguishable.
        Stars: optimal zero-work feedback rule (Fig.~\ref{fig:pnas_fig2}). 
        }
    \label{fig:dissipative_max}
\end{figure}

We perform gradient ascent of the performance function to find the Pareto frontier for this feedback scheme.  
Figure~\ref{fig:dissipative_max} shows that there is negligible difference between the Pareto frontier found for the nonnegative-work scheme and the dissipative scheme.
Thus, there is no added benefit to allowing moves with negative trap work, as they give up free energy that could have been extracted by an alternative move with zero trap work.
Thus, it is advantageous not to propose negative trap work moves at all if the objective is to maximize net output power.

\begin{figure}[!htbp]
    \centering  
    \includegraphics[clip, width=\linewidth]{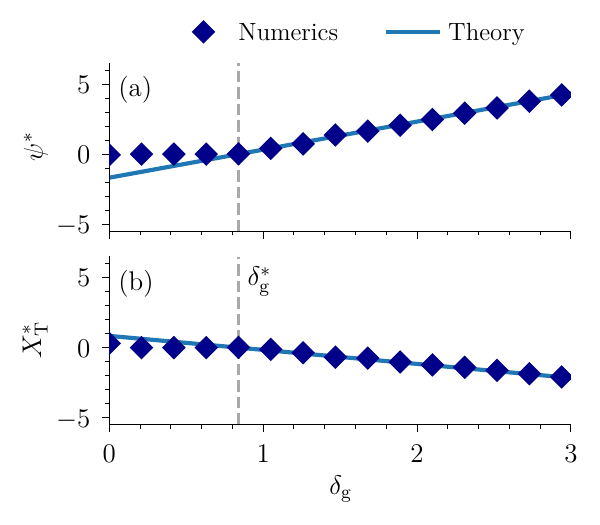}
    \caption{
        Dissipative-scheme parameters maximizing net output power.
        Optimal offset $\psi^{*}$ and threshold $\xT^{*}$ for dissipative feedback scheme with $\Gamma=1/2$ and fixed proportionality $\alpha^{*}=2$, as function of scaled effective mass $\dg$. 
        Diamonds: optimal parameters found by gradient ascent. 
        Solid lines: theory~\eqref{eq:strict_to_MD_rule_map}.
        Dashed vertical line: scaled effective mass $\dg^{*}\approx 0.84$ optimizing ZW performance.
        }
    \label{fig:diss_opt_net_output_params}
\end{figure}

To further emphasize this point, Fig.~\ref{fig:diss_opt_net_output_params} compares the optimal rules found for the dissipative scheme when the performance function equals the net output power (i.e., for $\Gamma = 1/2$). 
This shows even better agreement between the empirically found rules and the theoretical expectation~\eqref{eq:strict_to_MD_rule_map} than for the nonnegative-work scheme (Fig.~\ref{fig:opt_net_output_params_strict}).
In particular, rather than providing an upper bound on the feedback threshold $\xT$, for the dissipative feedback scheme our theoretical analysis~\eqref{eq:strict_to_MD_rule_map} reproduces exactly the empirically found rules when $\dg > \dg^{*}$. 
As the dissipative feedback scheme does not have the additional constraint imposed by requiring that all trap moves have an associated nonnegative trap work, the threshold $\xT$ is not a redundant variable. 
The excellent agreement then illustrates that the dissipative feedback schemes maximizing net output power do not propose any moves with negative trap work.

\section{Deriving steady-state distribution for no threshold}
\label{sec:ss_deriv}

Here we derive the steady-state distributions for the unconstrained feedback scheme used for calculations in Sec.~\ref{sec:full_extraction_scheme} of the main text.

From \eqref{eq:relative_pos_update} and \eqref{eq:relative_trap_update}, the particle's dynamics (for fixed trap center $\lk$) are an Ornstein-Uhlenbeck process with propagator
\begin{align}
    p\pr{\xkp|\xk,\lk} &= \mcal{N}\pr{\xkp; \mu_{\xkp|\xk}, \sigma_{\ts}^{2}}\ ,
\end{align}
with mean $\mu_{\xkp|\xk} = \xk\e^{-\ts} + \pr{1-\e^{-\ts}}\pr{\lk-\dg}$ and variance $\sigma_{\ts}^{2} = 1-\e^{-2\ts}$. 
The trap potential moves deterministically in response to a measurement of the particle's displacement $\xkpr$ from the trap center, with transition density
\begin{align}
    p\pr{\lkp|\xkp,\lk} = \delta\sr{\lkp-\pr{\lk + \alpha\xkpr + \psi}}\ ,
\end{align}
for Dirac delta function $\delta[\cdot]$.

A coordinate transformation from the lab frame to the trap frame with relative coordinates
\begin{subequations}
    \begin{align}
        \xkpr &\equiv \xkp-\lk\\
        \xkr &\equiv \xk-\lk
    \end{align}
\end{subequations}
gives
\begin{align}
    p_{1}\pr{\xkpr|\xkr} &= \mcal{N}\pr{\xkpr; \mu_{\xkpr|\xkr}, \sigma_{\ts}^{2}}\ ,
\end{align}
with mean $\mu_{\xkpr|\xkr} = \xkr\e^{-\ts}-\pr{1-\e^{-\ts}}\dg$.
Similarly, using $\xkrp = \xkp-\lkp$, the transition density of the trap is transformed in this frame to
\begin{align}
    p_{2}\pr{\xkrp|\xkpr} = \delta\sr{\xkrp + \pr{\alpha-1}\xkpr + \psi}.
\end{align}

The steady-state solutions for the distributions $\pi\pr{\xkpr}$ and $\pi\pr{\xkr}$ lead to the self-consistent equations
\begin{subequations}\label{eq:self_consistent_eqs}
    \begin{align}
        \pi\pr{\xkpr} &= \int\dd{v}\underbrace{\br{\int\dd{u}p_{1}\pr{\xkpr|u}p_{2}\pr{u|v}}}_{\equiv \, T\pr{\xkpr|v}}\pi\pr{v}\label{eq:self_consistent_eqs1}\\
        \tilde\pi\pr{\xkr} &= \int\dd{u}\underbrace{\br{\int\dd{v}p_{2}\pr{\xkr|v}p_{1}\pr{v|u}}}_{\equiv \, \tilde{T}\pr{\xkr|u}}\tilde\pi\pr{u}\ .\label{eq:self_consistent_eqs2}
    \end{align}
\end{subequations}

The propagator $T\pr{\xkpr|v}$ simplifies to
\begin{align}
    T\pr{\xkpr|v} &= \mcal{N}\pr{\xkpr;\mu_{\xkpr|v},\sigma_{\ts}^{2}}\ ,
\end{align}
with mean
\begin{align}
    \mu_{\xkpr|v} = -\sr{\pr{\alpha-1}v\e^{-\ts}+\pr{1-\e^{-\ts}}\dg + \psi\e^{-\ts}}\ .
\end{align} 
Similarly, the propagator $\tilde{T}\pr{\xkr|u}$ is
\begin{align}
    \tilde{T}\pr{\xkr|u} &= \mcal{N}\pr{\xkr;\mu_{\xkr|u},(1-\alpha)^2\,\sigma_{\ts}^{2}}\ ,
\end{align}
with mean
\begin{align}
    \mu_{\xkr|u} = -\pr{\alpha-1}u\e^{-\ts} + \pr{\alpha-1}\pr{1-\e^{-\ts}}\dg - \psi\ .
\end{align}

The steady-state distributions for both of these coordinates are Gaussian,
\begin{subequations}
    \begin{align}
        \pi\pr{\xkpr} &= \mcal{N}\pr{\xkpr;\mu_{\xkpr},\sigma_{\xkpr}^{2}}\label{eq:pi_xkpr}\\
        \tilde\pi\pr{\xkr} &= \mcal{N}\pr{\xkr;\mu_{\xkr},\sigma_{\xkr}^{2}}\ ,\label{eq:pi_xkr}
    \end{align}
\end{subequations}
each satisfying its corresponding self-consistent steady-state equation~\eqref{eq:self_consistent_eqs}.

Multiplying these equations by powers of the respective variables and integrating gives self-consistent equations for the mean and variance.
Multiplying \eqref{eq:self_consistent_eqs1} by $\xkpr$, using \eqref{eq:pi_xkpr}, and integrating yields
\begin{align}
    \mu_{\xkpr} &= -\pr{\alpha-1}\mu_{\xkpr}\e^{-\ts} + \pr{1-\e^{-\ts}}\dg - \e^{-\ts}\psi\ .\nonumber
\end{align}
Solving for the mean gives
\begin{align}
    \mu_{\xkpr} &= -\frac{\dg+(\psi-\dg)\e^{-\ts}}{1+(\alpha-1)\e^{-\ts}}.\label{eq:mean_dxkt}
\end{align}
The variance is computed similarly, yielding
\begin{align}
    \sigma_{\xkpr}^{2} &= \sigma_{\ts}^{2}+\sigma_{\xkpr}^{2}\pr{\alpha-1}^{2}\e^{-2\ts}\ . 
\end{align}
Solving for the variance gives
\begin{align}
    \sigma_{\xkpr}^{2} &= \frac{1-\e^{-2\ts}}{1-(\alpha-1)^{2}\e^{-2\ts}}\ .
\end{align}
These define the distribution~\eqref{eq:ss_natural_dxkt} in the main text. 

Performing the same procedure for $\xkr$, using \eqref{eq:self_consistent_eqs2} and \eqref{eq:pi_xkr}, gives mean 
\begin{align}
    \mu_{\xkr} &= \frac{\pr{1-\e^{-\ts}}\pr{\alpha-1}\dg -\psi}{1+(\alpha-1)\e^{-\ts}}\label{eq:mean_xkr} 
\end{align}
and variance
\begin{align}
    \sigma_{\xkr}^{2} &= \frac{\pr{1-\e^{-2\ts}} \pr{\alpha-1}^{2}}{1-(\alpha-1)^{2}\e^{-2\ts}}\ .
\end{align} 
These define the distribution~\eqref{eq:ss_natural_xkr} in the main text.

\section{Stochastic gradient ascent}
\label{sec:grad_ascent}

Here we describe the details of the stochastic gradient-ascent procedure that we use throughout this article.
Starting from an initial guess for the parameters, $\bm{\Phi}_{0} = \{\alpha,\psi,\xT\}_{0}$, the mean rate of free-energy change $\epg$ and the mean trap input power $\ept$ are estimated using \eqref{eq:nat_pgrav_def} and \eqref{eq:nat_ptrap_def} from an ensemble of 1000 particle-trap trajectories evolving according to the discrete-time Langevin equation~\eqref{eq:discrete_od_langevin} integrated from time $t=0$ to $t=1000$ (in units of relaxation time $\trelo$).
All simulations use $\fs=1000$ guided by our finding (in Sec.~\ref{subsubsec:performance_bounds_under_full_extraction}, Fig.~\ref{fig:pnas_fig}, 
and App.~\ref{sec:sampling_freq_variation}) that performance is generally maximized for high sampling frequencies.
A change in the parameters $\bm{\Phi}_{i+1} = \bm{\Phi}_{i} + \hat{\Sigma}\cdot\bm{\xi}$ is then proposed, with $\bm{\xi}$ a Gaussian random vector with zero mean and unit variance and $\Sigma_{ij} = \delta_{ij}\sigma_{i}$ a diagonal matrix with gradient-ascent parameters $\sigma_{i} = 1/2\ \forall\ i\in\{\alpha,\psi,\xT\}$. 
The proposed move is accepted if the performance $\mP$~\eqref{eq:gen_ratchet_benefit} improves; otherwise, it is rejected.
(This is equivalent to basin-hopping optimization~\cite{Wales1997} at zero temperature with objective function $-\mP$.)
The ascent proceeds for 800 proposed moves in parameter space.

In the unconstrained scheme, the initial guess is $\bm{\Phi}_{0} = \{\alpha=1,\psi=\xT=0\}$, while for the nonnegative-work and dissipative schemes the initial guess is the parameter set of the optimal zero-work feedback scheme, $\bm{\Phi}_{0} = \{\alpha=2,\psi=\xT=0\}$.

\section{Performance as a function of sampling frequency}
\label{sec:sampling_freq_variation}

\begin{figure}[!htbp]
    \centering 
    \includegraphics[clip, width=\linewidth]{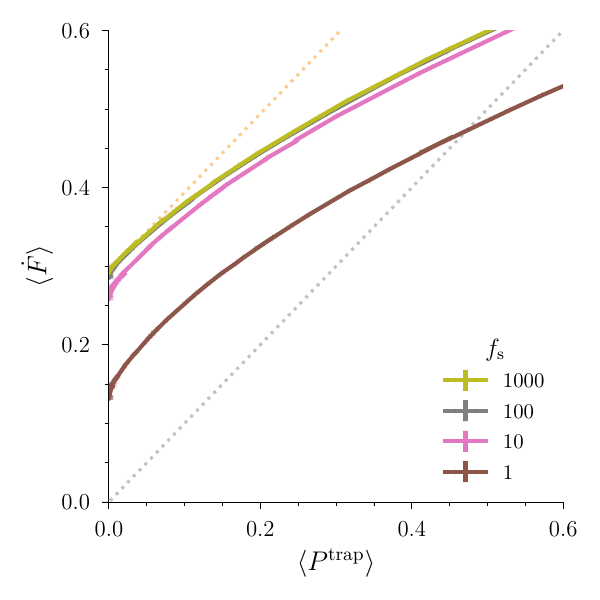}
    \caption{Pareto performance frontiers for the nonnegative-work feedback scheme for different values of the sampling frequency $\fs$ with scaled effective mass $\dg = 0.84$.
    Dotted orange line: net output power $\epn\approx 0.294$ of the optimal nonnegative-work feedback scheme.
    Dotted gray line: $\epn = 0$.
    }
    \label{fig:sampling_freq_var}
\end{figure}

Here we investigate how sampling frequency affects the Pareto performance frontiers of the optimal nonnegative-work feedback rules.
Figure~\ref{fig:sampling_freq_var} shows that the Pareto frontiers of the nonnegative-work feedback scheme associated with lower sampling frequencies are dominated (i.e., do worse in one objective without improving the other) by the frontiers associated with higher sampling frequencies. 
Lower sampling frequencies lead to both missed observations of the particle's first crossing of the threshold and threshold re-crossings, resulting in an overall reduced rate of energy extraction.

\section{Mapping dynamics under the zero-work scheme to dynamics under the nonnegative-work feedback scheme}
\label{sec:mapping}

Here we ``map'' the stochastic process imposed by the nonnegative-work feedback scheme to that under the optimal zero-work feedback rule, thereby yielding the choice of parameters~\eqref{eq:opt_net_output_params} in the main text.
We first use the coordinate
\begin{subequations}
    \begin{align}
        \xkprt &= \xkpr + \dg \\
        &= \xkp-\pr{\lk-\dg}\ , 
    \end{align}
\end{subequations}
to rewrite the optimal zero-work (ZW) feedback rule~\eqref{eq:opt_MD_rule} in the frame that is co-moving with the \emph{total} potential, obtaining
\begin{align}
    \pr{\dlk^{*}}^{\mrm{ZW}} = \Theta\pr{\xkprt-\dg^{*}}2\pr{\xkprt-\dg^{*}}\ ,\label{eq:opt_max_demon_rule_relative}
\end{align}
where $\dg^{*} \approx 0.84$. 
Now we compute the optimal nonnegative-work (NNW) scheme~\eqref{eq:strict_scheme} in this co-moving frame and obtain
\begin{align}
    \pr{\dlk^{*}}^{\mrm{NNW}} &= \Theta\pr{\xkprt-\dg-\xT^{*}}\sr{\alpha^*\pr{\xkprt-\dg}+\psi^{*}}\ . \label{eq:opt_strict_scheme_relative}
\end{align}
The Heaviside function in \eqref{eq:strict_scheme2} imposing positive trap input work is implicitly accounted for in the allowed choices of the parameters.
Feedback rules \eqref{eq:opt_max_demon_rule_relative} and \eqref{eq:opt_strict_scheme_relative} would be identical for
\begin{subequations}\label{eq:strict_to_MD_rule_map}
    \begin{align}
        \alpha^{*} &= 2,\\
        \psi^{*} &= 2\pr{\dg-\dg^{*}},\\
        \xT^{*} &= \dg^{*}-\dg = -\frac{1}{2}\psi^{*}\ .
    \end{align}
\end{subequations}

Choosing the parameters as above amounts to fixing different starting position $\xR$ and threshold position $\xT$. 
Consequently, this choice---similar to what was seen in Sec.~\ref{subsubsec:maxwell_demon_feedback_scheme}---affects its mean first-passage time, i.e., the time it takes for the particle to traverse the distance between its starting position and the threshold.
In particular, this choice shifts the respective optimal start and threshold locations $\xR$ and $\xT$ of the optimal zero-work feedback rule, yielding
\begin{subequations}
    \begin{align}
        \xR &= \xR^{*} + (\dg - \dg^{*}) \\
        \xT &= \xT^{*} - (\dg - \dg^{*}) \ .
    \end{align}
\end{subequations}
The net output work associated with the optimal parameters is then
\begin{subequations}
    \begin{align}
        \wn&\pr{\dg,\xR,\xT} \nonumber \\
        &= \wg\pr{\dg,\xR, \xT} - \wt\pr{\xR, \xT}\\
        &= \dg\pr{\xT+\xR} - \frac{1}{2}\pr{\xR^{2}-\xT^{2}}\\
        &= \dg^{*}\pr{\xT^{*}+\xR^{*}} - \frac{1}{2}
        \pr{\xR^{2}-\xT^{2}}\\
        &= \wn\pr{\dg^{*},\xR^{*},\xT^{*}}\ .
    \end{align}
\end{subequations}
The mean first-passage time~\eqref{eq:MFPT}, using $x' = y - \dg + \dg^{*}$ and $x'' = y'-\dg+\dg^{*}$, yields~\cite{Hanggi1990,Pontryagin1933} 
\begin{align}
    \tmfp\pr{\xR,\xT} = \mkern-18mu\int\limits_{-\sr{\xR-(\dg-\dg^{*})}}^{\xT+(\dg-\dg^{*})}\mkern-18mu\dd{y}\e^{V^{*}(y)}\int_{-\infty}^{y}\dd{y'}\e^{-V^{*}(y')}\ ,
\end{align}
where $V^{*}(x) = \frac{1}{2}x^{2} + \dg^{*}x$ is the total potential associated with the optimal scaled effective mass $\dg^{*}$. 
Computing the net output power, 
\begin{align}
    \epn = \frac{\wg\pr{\xR,\xT}-\wt\pr{\xR,\xT}}{\tmfp\pr{\xR,\xT}}\ ,
\end{align}
and taking the limit as the optimal reset and threshold locations approach one another (as in Sec.~\ref{subsubsec:maxwell_demon_feedback_scheme}) recovers 
\begin{align}
    \lim_{\xR\to(\dg-\dg^{*})}\, \lim_{\xT\to(\dg^{*}-\dg)}&\epn \\
    &\hspace{-1.5cm}= \sqrt{\frac{2}{\pi}}\dg^{*}\e^{-(\dg^{*})^{2}/2}\sr{1+\erf\pr{\frac{\dg^{*}}{\sqrt{2}}}}^{-1}\ .\nonumber
\end{align}
Therefore, this choice of parameters for the nonnegative-work feedback scheme does return the maximum value of the net output power~\eqref{eq:pow_theory_max} obtained in the optimal zero-work rule, for any $\dg > \dg^{*}$. 

{\color{black} 
\section{Optimal nonnegative-work feedback scheme has the same net output work distribution as the zero-work scheme}
\label{sec:work_distributions}

We show in Fig.~\ref{fig:opt_strict_scheme_dg_var} that the stationary distribution of the nonnegative-work scheme under the choice of 
parameters~\eqref{eq:opt_net_output_params} is equivalent to that of the optimal zero-work scheme~\eqref{eq:opt_MD_rule} with $\dg = \dg^{*}$. 
Here, we show that this result implies the net output power distribution must also be the same for the two feedback schemes. 

To calculate the distribution for the net output work for one step of the dynamics, we require the joint distribution of the relative coordinates
\begin{align}
    p\pr{\xkpr, \xkr} &= \pi^{\mrm{NNW}}\pr{\xkpr; \dg}\nonumber\\
    &\hspace{-1.5cm}\times\Bigg\{\Theta\pr{-\sr{\xkpr-\xT^{*}}}\delta\pr{\xkr-\xkrp}\\
    &\hspace{-0.9cm}+ \Theta\pr{\xkpr-\xT^{*}}\delta\pr{\xkr-\sr{\xkpr-\pr{\alpha^{*}\xkpr+\psi^{*}}}}\Bigg\}\nonumber\\
    &\hspace{-1.2cm} = \pi^{\mrm{NNW}}\pr{\xkpr;\dg}\Bigg\{\Theta\pr{-\sr{\xkpr+\dg-\dg^{*}}}\delta\pr{\xkr-\xkrp}\nonumber\\
    &\hspace{-1cm}+ \Theta\pr{\xkpr+\dg-\dg^{*}}\delta\pr{\xkr-\sr{-\xkpr-2\pr{\dg-\dg^{*}}}}\Bigg\}\ ,
\end{align}
where the superscript NNW on the stationary distribution $\pi^{\mrm{NNW}}$ emphasizes that this is the stationary distribution of the nonnegative-work scheme. 
The parameters $\alpha^{*}$, $\psi^{*}$ and $\xT^{*}$ are found in~\eqref{eq:opt_net_output_params}.
We transform this distribution from the trap-potential frame to the total-potential frame with coordinates $\xkprt\equiv \xkpr + \dg$ and $\xkrt\equiv r_{k}+\dg$:
\begin{subequations}\label{eq:s_joint_distribution}
    \begin{align}
        p\pr{\xkprt, \xkrt} &= \int\dd{\xkprt}\dd{\xkrt}\, \Bigg\{p\pr{\xkprt, \xkrt}\nonumber\\
        &\hspace{-2.1cm}\times\delta\pr{\xkprt-r_{k}^{+}-\dg}\delta\pr{\xkrt-r_{k}-\dg}\Bigg\}\\
        &\hspace{-1.2cm}=\pi^{\mrm{NNW}}\pr{\xkprt-\dg; \dg}\nonumber\\
        &\hspace{-2.5cm}\times\Bigg\{\Theta\pr{-\xkprt+\dg^{*}}\delta\pr{\xkrt-\xkprt}\\
        &\hspace{-1.2cm}+\Theta\pr{\xkprt-\dg^{*}}\delta\pr{\xkrt + \xkprt-2\dg^{*}}\Bigg\}\nonumber\ .
    \end{align}
\end{subequations}
The dependence of this expression on the scaled mass $\dg$ is contained entirely in the stationary distribution $\pi^{\mrm{NNW}}$; the sum in the braces is entirely independent of $\dg$. 
Figure~\ref{fig:opt_strict_scheme_dg_var}b shows empirically that
\begin{subequations}
    \begin{align}
        \pi^{\mrm{NNW}}\pr{\xkpr - \dg; \dg} &= \pi^{\mrm{ZW}}\pr{\xkpr - \dg^{*}; \dg^{*}}\label{eq:distr_equality}\\
        \pi^{\mrm{NNW}}\pr{\xkpr; \dg} &= \pi^{\mrm{ZW}}\pr{\xkpr + \dg - \dg^{*}; \dg^{*}}\\
        \pi^{\mrm{NNW}}\pr{\xkprt-\dg; \dg} &= \pi^{\mrm{ZW}}\pr{\xkprt - \dg^{*}; \dg^{*}}\ ,
    \end{align}
\end{subequations}
for the steady-state distribution $\pi^{\mrm{ZW}}$ of the zero-work scheme.
Substituting this result into \eqref{eq:s_joint_distribution} gives
\begin{align}
    p\pr{\xkprt,\xkrt} &= \pi^{\mrm{ZW}}\pr{\xkprt-\dg^{*}; \dg^{*}}\nonumber\\
        &\hspace{-1.8cm}\times\Bigg\{\Theta\pr{-\xkprt+\dg^{*}}\delta\pr{\xkrt-\xkprt}\\
        &\hspace{-1.2cm}+\Theta\pr{\xkprt-\dg^{*}}\delta\pr{\xkrt + \xkprt-2\dg^{*}}\Bigg\}\nonumber\ ,\label{eq:s_joint_distribution_indep}
\end{align}
which is now completely independent of the scaled mass $\dg$.

The net output power is, in the frame of the total potential, 
\begin{align}
    \pn = \frac{1}{2\ts}\sr{\pr{\xkprt}^{2} - \pr{\xkrt}^{2}}\ .
\end{align}
Hence we compute the stationary distribution over net output power as
\begin{align}
    p\pr{\pn} &= \int\dd{\xkprt}\dd{\xkrt}p\pr{\xkprt, \xkrt}\\
    &\times\delta\pr{\pn - \frac{1}{2\ts}\sr{\pr{\xkprt}^{2} - \pr{\xkrt}^{2}}}\ ,\nonumber
\end{align}
independent of $\dg$.

We have thus shown that by the empirical equality of distributions~\eqref{eq:distr_equality} shown in Fig.~\ref{fig:opt_strict_scheme_dg_var}b, the nonnegative-work scheme using the parameters~\eqref{eq:opt_net_output_params} achieves the same net output power \emph{distribution} per measurement as that of the optimal zero-work scheme~\eqref{eq:opt_MD_rule} for $\dg^{*}$.

For completeness, Fig.~\ref{fig:work_distributions} shows the simulated distributions of the net output power $\pn$, the rate of free-energy change $\pg$, and the trap input power $\pt$.
Figure~\ref{fig:work_distributions}a shows that the net output power distributions for the optimal zero-work scheme and the optimal nonnegative-work scheme closely match, in agreement with the above analysis. 
Figures~\ref{fig:work_distributions}b,c show the distributions of the rate of free-energy change $\pg$ and trap input power $\pt$, respectively.
While the net output power distribution is preserved in moving from the non-optimal zero-work rule to the optimal nonnegative-work feedback rule for scaled masses $\dg \gtrsim 0.84$, 
the distributions for $\pg$ and $\pt$ are not.
In general, the variance of these distributions increases as the scaled mass $\dg$ increases, as well as in going from a non-optimal zero-work feedback rule to the 
optimal nonnegative-work feedback rule thereby allowing for the input of energy.

\begin{figure}
    \centering
    \includegraphics[clip, width=\linewidth]{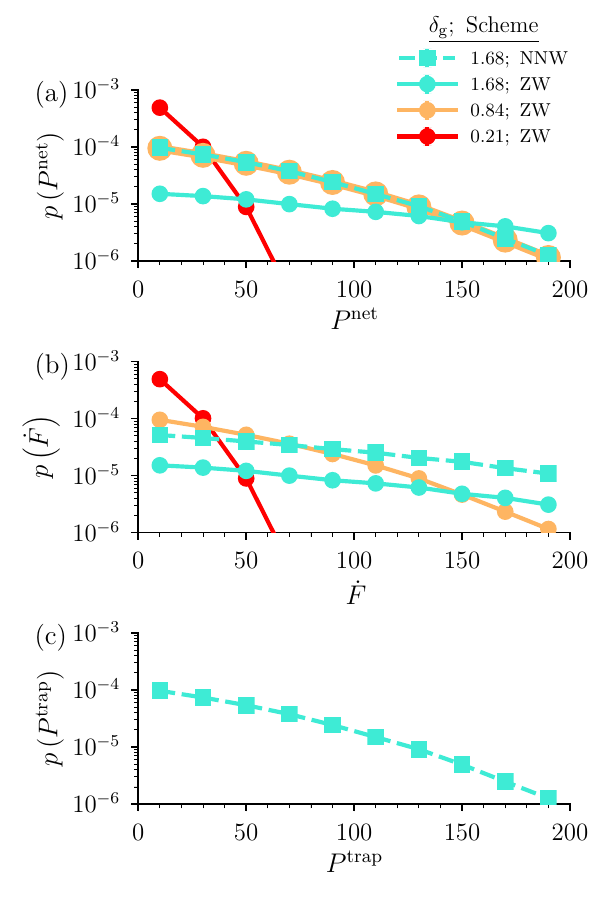}
    \caption{\color{black} Empirical distributions of power for various feedback schemes.   
    (a) Net output power, (b) rate of free-energy change, and (c) trap input power
    (where, by construction, the zero-work feedback rules deterministically produce $\pt = 0$ for every measurement). 
    For clarity, we omit the probability of zero power. 
    Different colors denote different $\dg$. Squares and solid connecting curves: optimal zero-work (ZW) feedback rule~\eqref{eq:opt_MD_rule}. 
    Points and dashed connecting curves: optimal nonnegative work (NNW) feedback rule~\eqref{eq:opt_net_output_params}. 
    Each histogram of 10 bins is constructed from a long ($t = 10^{4}\trelo)$ trajectory with sampling frequency $\fs = 1000$. 
    }
    \label{fig:work_distributions}
\end{figure}

}

\FloatBarrier
\newpage


\begin{thebibliography}{36}%
	\makeatletter
	\providecommand \@ifxundefined [1]{%
		\@ifx{#1\undefined}
	}%
	\providecommand \@ifnum [1]{%
		\ifnum #1\expandafter \@firstoftwo
		\else \expandafter \@secondoftwo
		\fi
	}%
	\providecommand \@ifx [1]{%
		\ifx #1\expandafter \@firstoftwo
		\else \expandafter \@secondoftwo
		\fi
	}%
	\providecommand \natexlab [1]{#1}%
	\providecommand \enquote  [1]{``#1''}%
	\providecommand \bibnamefont  [1]{#1}%
	\providecommand \bibfnamefont [1]{#1}%
	\providecommand \citenamefont [1]{#1}%
	\providecommand \href@noop [0]{\@secondoftwo}%
	\providecommand \href [0]{\begingroup \@sanitize@url \@href}%
	\providecommand \@href[1]{\@@startlink{#1}\@@href}%
	\providecommand \@@href[1]{\endgroup#1\@@endlink}%
	\providecommand \@sanitize@url [0]{\catcode `\\12\catcode `\$12\catcode
		`\&12\catcode `\#12\catcode `\^12\catcode `\_12\catcode `\%12\relax}%
	\providecommand \@@startlink[1]{}%
	\providecommand \@@endlink[0]{}%
	\providecommand \url  [0]{\begingroup\@sanitize@url \@url }%
	\providecommand \@url [1]{\endgroup\@href {#1}{\urlprefix }}%
	\providecommand \urlprefix  [0]{URL }%
	\providecommand \Eprint [0]{\href }%
	\providecommand \doibase [0]{https://doi.org/}%
	\providecommand \selectlanguage [0]{\@gobble}%
	\providecommand \bibinfo  [0]{\@secondoftwo}%
	\providecommand \bibfield  [0]{\@secondoftwo}%
	\providecommand \translation [1]{[#1]}%
	\providecommand \BibitemOpen [0]{}%
	\providecommand \bibitemStop [0]{}%
	\providecommand \bibitemNoStop [0]{.\EOS\space}%
	\providecommand \EOS [0]{\spacefactor3000\relax}%
	\providecommand \BibitemShut  [1]{\csname bibitem#1\endcsname}%
	\let\auto@bib@innerbib\@empty
	\bibitem [{\citenamefont {Maxwell}(1878)}]{Maxwell1878}%
	\BibitemOpen
	\bibfield  {author} {\bibinfo {author} {\bibfnamefont {J.~C.}\ \bibnamefont
			{Maxwell}},\ }\bibfield  {title} {\bibinfo {title} {{Tait's
				``Thermodynamics''}},\ }\href {https://doi.org/10.1038/017257a0} {\bibfield
		{journal} {\bibinfo  {journal} {Nature}\ }\textbf {\bibinfo {volume} {17}},\
		\bibinfo {pages} {278} (\bibinfo {year} {1878})}\BibitemShut {NoStop}%
	\bibitem [{\citenamefont {Szil\'ard}(1964)}]{Szilard1964}%
	\BibitemOpen
	\bibfield  {author} {\bibinfo {author} {\bibfnamefont {L.}~\bibnamefont
			{Szil\'ard}},\ }\bibfield  {title} {\bibinfo {title} {On the decrease of
			entropy in a thermodynamic system by the intervention of intelligent
			beings},\ }\href {https://doi.org/https://doi.org/10.1002/bs.3830090402}
	{\bibfield  {journal} {\bibinfo  {journal} {Behav. Sci.}\ }\textbf {\bibinfo
			{volume} {9}},\ \bibinfo {pages} {301} (\bibinfo {year} {1964})}\BibitemShut
	{NoStop}%
	\bibitem [{\citenamefont {Paneru}\ and\ \citenamefont
		{Kyu~Pak}(2020)}]{Paneru2020}%
	\BibitemOpen
	\bibfield  {author} {\bibinfo {author} {\bibfnamefont {G.}~\bibnamefont
			{Paneru}}\ and\ \bibinfo {author} {\bibfnamefont {H.}~\bibnamefont
			{Kyu~Pak}},\ }\bibfield  {title} {\bibinfo {title} {Colloidal engines for
			innovative tests of information thermodynamics},\ }\href
	{https://doi.org/10.1080/23746149.2020.1823880} {\bibfield  {journal}
		{\bibinfo  {journal} {Adv. Phys.:X}\ }\textbf {\bibinfo {volume} {5}},\
		\bibinfo {pages} {1823880} (\bibinfo {year} {2020})}\BibitemShut {NoStop}%
	\bibitem [{\citenamefont {Leff}\ and\ \citenamefont
		{Rex}(2002)}]{leff_book2002}%
	\BibitemOpen
	\bibfield  {author} {\bibinfo {author} {\bibfnamefont {H.}~\bibnamefont
			{Leff}}\ and\ \bibinfo {author} {\bibfnamefont {A.}~\bibnamefont {Rex}},\
	}\href@noop {} {\emph {\bibinfo {title} {Maxwell's Demon 2: Entropy,
				Classical and Quantum Information, Computing}}}\ (\bibinfo  {publisher} {CRC
		Press},\ \bibinfo {year} {2002})\BibitemShut {NoStop}%
	\bibitem [{\citenamefont {Bennett}(1982)}]{Bennett1982}%
	\BibitemOpen
	\bibfield  {author} {\bibinfo {author} {\bibfnamefont {C.~H.}\ \bibnamefont
			{Bennett}},\ }\bibfield  {title} {\bibinfo {title} {The thermodynamics of
			computation---a review},\ }\href {https://doi.org/10.1007/BF02084158}
	{\bibfield  {journal} {\bibinfo  {journal} {Int. J. Theor. Phys.}\ }\textbf
		{\bibinfo {volume} {21}},\ \bibinfo {pages} {905} (\bibinfo {year}
		{1982})}\BibitemShut {NoStop}%
	\bibitem [{\citenamefont {Toyabe}\ \emph {et~al.}(2010)\citenamefont {Toyabe},
		\citenamefont {Sagawa}, \citenamefont {Ueda}, \citenamefont {Muneyuki},\ and\
		\citenamefont {Sano}}]{Toyabe2010}%
	\BibitemOpen
	\bibfield  {author} {\bibinfo {author} {\bibfnamefont {S.}~\bibnamefont
			{Toyabe}}, \bibinfo {author} {\bibfnamefont {T.}~\bibnamefont {Sagawa}},
		\bibinfo {author} {\bibfnamefont {M.}~\bibnamefont {Ueda}}, \bibinfo {author}
		{\bibfnamefont {E.}~\bibnamefont {Muneyuki}},\ and\ \bibinfo {author}
		{\bibfnamefont {M.}~\bibnamefont {Sano}},\ }\bibfield  {title} {\bibinfo
		{title} {{Experimental demonstration of information-to-energy conversion and
				validation of the generalized Jarzynski equality}},\ }\href
	{https://doi.org/10.1038/nphys1821} {\bibfield  {journal} {\bibinfo
			{journal} {Nat. Phys.}\ }\textbf {\bibinfo {volume} {6}},\ \bibinfo {pages}
		{988} (\bibinfo {year} {2010})}\BibitemShut {NoStop}%
	\bibitem [{\citenamefont {Lee}\ \emph {et~al.}(2018)\citenamefont {Lee},
		\citenamefont {Um}, \citenamefont {Paneru},\ and\ \citenamefont
		{Pak}}]{Lee2018}%
	\BibitemOpen
	\bibfield  {author} {\bibinfo {author} {\bibfnamefont {D.~Y.}\ \bibnamefont
			{Lee}}, \bibinfo {author} {\bibfnamefont {J.}~\bibnamefont {Um}}, \bibinfo
		{author} {\bibfnamefont {G.}~\bibnamefont {Paneru}},\ and\ \bibinfo {author}
		{\bibfnamefont {H.~K.}\ \bibnamefont {Pak}},\ }\bibfield  {title} {\bibinfo
		{title} {{An experimentally-achieved information-driven Brownian motor shows
				maximum power at the relaxation time}},\ }\href
	{https://doi.org/10.1038/s41598-018-30495-6} {\bibfield  {journal} {\bibinfo
			{journal} {Sci. Rep.}\ }\textbf {\bibinfo {volume} {8}},\ \bibinfo {pages}
		{12121} (\bibinfo {year} {2018})}\BibitemShut {NoStop}%
	\bibitem [{\citenamefont {Admon}\ \emph {et~al.}(2018)\citenamefont {Admon},
		\citenamefont {Rahav},\ and\ \citenamefont {Roichman}}]{admon2018}%
	\BibitemOpen
	\bibfield  {author} {\bibinfo {author} {\bibfnamefont {T.}~\bibnamefont
			{Admon}}, \bibinfo {author} {\bibfnamefont {S.}~\bibnamefont {Rahav}},\ and\
		\bibinfo {author} {\bibfnamefont {Y.}~\bibnamefont {Roichman}},\ }\bibfield
	{title} {\bibinfo {title} {{Experimental Realization of an Information
				Machine with Tunable Temporal Correlations}},\ }\href
	{https://doi.org/10.1103/PhysRevLett.121.180601} {\bibfield  {journal}
		{\bibinfo  {journal} {Phys. Rev. Lett.}\ }\textbf {\bibinfo {volume} {121}},\
		\bibinfo {pages} {180601} (\bibinfo {year} {2018})}\BibitemShut {NoStop}%
	\bibitem [{\citenamefont {Ribezzi-Crivellari}\ and\ \citenamefont
		{Ritort}(2019)}]{Ribezzi-Crivellari2019}%
	\BibitemOpen
	\bibfield  {author} {\bibinfo {author} {\bibfnamefont {M.}~\bibnamefont
			{Ribezzi-Crivellari}}\ and\ \bibinfo {author} {\bibfnamefont
			{F.}~\bibnamefont {Ritort}},\ }\bibfield  {title} {\bibinfo {title} {{Large
				work extraction and the Landauer limit in a continuous Maxwell demon}},\
	}\href {https://doi.org/10.1038/s41567-019-0481-0} {\bibfield  {journal}
		{\bibinfo  {journal} {Nat. Phys.}\ }\textbf {\bibinfo {volume} {15}},\
		\bibinfo {pages} {660} (\bibinfo {year} {2019})}\BibitemShut {NoStop}%
	\bibitem [{\citenamefont {Paneru}\ \emph {et~al.}(2018)\citenamefont {Paneru},
		\citenamefont {Lee}, \citenamefont {Park}, \citenamefont {Park},
		\citenamefont {Noh},\ and\ \citenamefont {Pak}}]{Paneru2018b}%
	\BibitemOpen
	\bibfield  {author} {\bibinfo {author} {\bibfnamefont {G.}~\bibnamefont
			{Paneru}}, \bibinfo {author} {\bibfnamefont {D.~Y.}\ \bibnamefont {Lee}},
		\bibinfo {author} {\bibfnamefont {J.-M.}\ \bibnamefont {Park}}, \bibinfo
		{author} {\bibfnamefont {J.~T.}\ \bibnamefont {Park}}, \bibinfo {author}
		{\bibfnamefont {J.~D.}\ \bibnamefont {Noh}},\ and\ \bibinfo {author}
		{\bibfnamefont {H.~K.}\ \bibnamefont {Pak}},\ }\bibfield  {title} {\bibinfo
		{title} {Optimal tuning of a brownian information engine operating in a
			nonequilibrium steady state},\ }\href
	{https://doi.org/10.1103/PhysRevE.98.052119} {\bibfield  {journal} {\bibinfo
			{journal} {Phys. Rev. E}\ }\textbf {\bibinfo {volume} {98}},\ \bibinfo
		{pages} {052119} (\bibinfo {year} {2018})}\BibitemShut {NoStop}%
	\bibitem [{\citenamefont {Cao}\ and\ \citenamefont {Feito}(2009)}]{Cao2009}%
	\BibitemOpen
	\bibfield  {author} {\bibinfo {author} {\bibfnamefont {F.~J.}\ \bibnamefont
			{Cao}}\ and\ \bibinfo {author} {\bibfnamefont {M.}~\bibnamefont {Feito}},\
	}\bibfield  {title} {\bibinfo {title} {Thermodynamics of feedback controlled
			systems},\ }\href {https://doi.org/10.1103/PhysRevE.79.041118} {\bibfield
		{journal} {\bibinfo  {journal} {Phys. Rev. E}\ }\textbf {\bibinfo {volume}
			{79}},\ \bibinfo {pages} {041118} (\bibinfo {year} {2009})}\BibitemShut
	{NoStop}%
	\bibitem [{\citenamefont {Jarzynski}(2011)}]{Jarzynski2011}%
	\BibitemOpen
	\bibfield  {author} {\bibinfo {author} {\bibfnamefont {C.}~\bibnamefont
			{Jarzynski}},\ }\bibfield  {title} {\bibinfo {title} {Equalities and
			inequalities: Irreversibility and the second law of thermodynamics at the
			nanoscale},\ }\href {https://doi.org/10.1088/1742-5468/2007/09/P09012}
	{\bibfield  {journal} {\bibinfo  {journal} {Annu. Rev. Condens. Matter}\
		}\textbf {\bibinfo {volume} {2}},\ \bibinfo {pages} {329} (\bibinfo {year}
		{2011})}\BibitemShut {NoStop}%
	\bibitem [{\citenamefont {Seifert}(2012)}]{seifert_2012}%
	\BibitemOpen
	\bibfield  {author} {\bibinfo {author} {\bibfnamefont {U.}~\bibnamefont
			{Seifert}},\ }\bibfield  {title} {\bibinfo {title} {Stochastic
			thermodynamics, fluctuation theorems and molecular machines},\ }\href
	{https://doi.org/10.1088/0034-4885/75/12/126001} {\bibfield  {journal}
		{\bibinfo  {journal} {Rep. Prog. Phys.}\ }\textbf {\bibinfo {volume} {75}},\
		\bibinfo {pages} {126001} (\bibinfo {year} {2012})}\BibitemShut {NoStop}%
	\bibitem [{\citenamefont {Parrondo}\ \emph {et~al.}(2015)\citenamefont
		{Parrondo}, \citenamefont {Horowitz},\ and\ \citenamefont
		{Sagawa}}]{Parrondo2015a}%
	\BibitemOpen
	\bibfield  {author} {\bibinfo {author} {\bibfnamefont {J.~M.~R.}\
			\bibnamefont {Parrondo}}, \bibinfo {author} {\bibfnamefont {J.~M.}\
			\bibnamefont {Horowitz}},\ and\ \bibinfo {author} {\bibfnamefont
			{T.}~\bibnamefont {Sagawa}},\ }\bibfield  {title} {\bibinfo {title}
		{{Thermodynamics of information}},\ }\href
	{https://doi.org/10.1038/nphys3230} {\bibfield  {journal} {\bibinfo
			{journal} {Nat. Phys.}\ }\textbf {\bibinfo {volume} {11}},\ \bibinfo {pages}
		{131} (\bibinfo {year} {2015})}\BibitemShut {NoStop}%
	\bibitem [{\citenamefont {Saha}\ \emph {et~al.}(2021)\citenamefont {Saha},
		\citenamefont {Lucero}, \citenamefont {Ehrich}, \citenamefont {Sivak},\ and\
		\citenamefont {Bechhoefer}}]{Saha2021}%
	\BibitemOpen
	\bibfield  {author} {\bibinfo {author} {\bibfnamefont {T.~K.}\ \bibnamefont
			{Saha}}, \bibinfo {author} {\bibfnamefont {J.~N.~E.}\ \bibnamefont {Lucero}},
		\bibinfo {author} {\bibfnamefont {J.}~\bibnamefont {Ehrich}}, \bibinfo
		{author} {\bibfnamefont {D.~A.}\ \bibnamefont {Sivak}},\ and\ \bibinfo
		{author} {\bibfnamefont {J.}~\bibnamefont {Bechhoefer}},\ }\bibfield  {title}
	{\bibinfo {title} {Maximizing power and velocity of an information engine},\
	}\href {https://doi.org/10.1073/pnas.2023356118} {\bibfield  {journal}
		{\bibinfo  {journal} {Proc. Natl. Acad. Sci. USA}\ }\textbf {\bibinfo
			{volume} {118}},\  (\bibinfo {year} {2021})}\BibitemShut {NoStop}%
	\bibitem [{\citenamefont {Gardiner}(2009)}]{gardiner_book2009}%
	\BibitemOpen
	\bibfield  {author} {\bibinfo {author} {\bibfnamefont {C.}~\bibnamefont
			{Gardiner}},\ }\href@noop {} {\emph {\bibinfo {title} {Stochastic Methods: A
				Handbook for the Natural and Social Sciences}}},\ Springer Series in
	Synergetics\ (\bibinfo  {publisher} {Springer Berlin Heidelberg},\ \bibinfo
	{year} {2009})\BibitemShut {NoStop}%
	\bibitem [{Note1()}]{Note1}%
	\BibitemOpen
	\bibinfo {note} {In experimental realizations, there is typically feedback
		latency -- a delay between the measurement time and the response time. This
		arises from the time required to transfer the measurement to the core
		processor of the hardware, compute the response, and communicate to the
		device that moves the trap. A typical feedback latency is $t_{\protect
			\mathrm {s}}$, which leads to small performance reductions that are neglected
		here.}\BibitemShut {Stop}%
	\bibitem [{Note2()}]{Note2}%
	\BibitemOpen
	\bibinfo {note} {In general, the particle will be in a nonequilibrium state;
		however, we consider the equilibrium free energy since we implicitly assume
		that the particle relaxes to equilibrium at the end of the process.
		Therefore, we can understand the gain in gravitational potential energy
		through the feedback operation as a change in free energy which can be
		related to the work input. As the equilibrium distribution is a function only
		of the difference $x-\lambda $, the equilibrium distribution at the end of
		the process (which the system would eventually relax to if nothing else were
		to happen) would be unchanged. As a result, the free energy changes according
		to \protect \textup {\hbox {\mathsurround \z@ \protect \normalfont
				(\ignorespaces \ref {eq:free_energy_change}\unskip \@@italiccorr
				)}}.}\BibitemShut {Stop}%
	\bibitem [{\citenamefont {Sagawa}\ and\ \citenamefont
		{Ueda}(2010)}]{Sagawa2010}%
	\BibitemOpen
	\bibfield  {author} {\bibinfo {author} {\bibfnamefont {T.}~\bibnamefont
			{Sagawa}}\ and\ \bibinfo {author} {\bibfnamefont {M.}~\bibnamefont {Ueda}},\
	}\bibfield  {title} {\bibinfo {title} {{Generalized Jarzynski Equality under
				Nonequilibrium Feedback Control}},\ }\href
	{https://doi.org/10.1103/PhysRevLett.104.090602} {\bibfield  {journal}
		{\bibinfo  {journal} {Phys. Rev. Lett.}\ }\textbf {\bibinfo {volume} {104}},\
		\bibinfo {pages} {090602} (\bibinfo {year} {2010})}\BibitemShut {NoStop}%
	\bibitem [{\citenamefont {Horowitz}\ and\ \citenamefont
		{Vaikuntanathan}(2010)}]{Horowitz2010}%
	\BibitemOpen
	\bibfield  {author} {\bibinfo {author} {\bibfnamefont {J.~M.}\ \bibnamefont
			{Horowitz}}\ and\ \bibinfo {author} {\bibfnamefont {S.}~\bibnamefont
			{Vaikuntanathan}},\ }\bibfield  {title} {\bibinfo {title} {Nonequilibrium
			detailed fluctuation theorem for repeated discrete feedback},\ }\href
	{https://doi.org/10.1103/PhysRevE.82.061120} {\bibfield  {journal} {\bibinfo
			{journal} {Phys. Rev. E}\ }\textbf {\bibinfo {volume} {82}},\ \bibinfo
		{pages} {061120} (\bibinfo {year} {2010})}\BibitemShut {NoStop}%
	\bibitem [{\citenamefont {Sagawa}\ and\ \citenamefont
		{Ueda}(2012)}]{Sagawa2012}%
	\BibitemOpen
	\bibfield  {author} {\bibinfo {author} {\bibfnamefont {T.}~\bibnamefont
			{Sagawa}}\ and\ \bibinfo {author} {\bibfnamefont {M.}~\bibnamefont {Ueda}},\
	}\bibfield  {title} {\bibinfo {title} {Nonequilibrium thermodynamics of
			feedback control},\ }\href {https://doi.org/10.1103/PhysRevE.85.021104}
	{\bibfield  {journal} {\bibinfo  {journal} {Phys. Rev. E}\ }\textbf {\bibinfo
			{volume} {85}},\ \bibinfo {pages} {021104} (\bibinfo {year}
		{2012})}\BibitemShut {NoStop}%
	\bibitem [{\citenamefont {Granger}\ \emph {et~al.}(2016)\citenamefont
		{Granger}, \citenamefont {Dinis}, \citenamefont {Horowitz},\ and\
		\citenamefont {Parrondo}}]{Granger2016}%
	\BibitemOpen
	\bibfield  {author} {\bibinfo {author} {\bibfnamefont {L.}~\bibnamefont
			{Granger}}, \bibinfo {author} {\bibfnamefont {L.}~\bibnamefont {Dinis}},
		\bibinfo {author} {\bibfnamefont {J.~M.}\ \bibnamefont {Horowitz}},\ and\
		\bibinfo {author} {\bibfnamefont {J.~M.~R.}\ \bibnamefont {Parrondo}},\
	}\bibfield  {title} {\bibinfo {title} {Reversible feedback confinement},\
	}\href {https://doi.org/10.1209/0295-5075/115/50007} {\bibfield  {journal}
		{\bibinfo  {journal} {Europhys. Lett.}\ }\textbf {\bibinfo {volume} {115}},\
		\bibinfo {pages} {50007} (\bibinfo {year} {2016})}\BibitemShut {NoStop}%
	\bibitem [{\citenamefont {Dinis}\ and\ \citenamefont
		{Parrondo}(2021)}]{Dinis2021}%
	\BibitemOpen
	\bibfield  {author} {\bibinfo {author} {\bibfnamefont {L.}~\bibnamefont
			{Dinis}}\ and\ \bibinfo {author} {\bibfnamefont {J.~M.~R.}\ \bibnamefont
			{Parrondo}},\ }\bibfield  {title} {\bibinfo {title} {{Extracting Work
				Optimally with Imprecise Measurements}},\ }\href
	{https://doi.org/10.3390/e23010008} {\bibfield  {journal} {\bibinfo
			{journal} {Entropy}\ }\textbf {\bibinfo {volume} {23}},\ \bibinfo {pages} {8}
		(\bibinfo {year} {2021})}\BibitemShut {NoStop}%
	\bibitem [{\citenamefont {Solon}\ and\ \citenamefont
		{Horowitz}(2018)}]{Solon2018}%
	\BibitemOpen
	\bibfield  {author} {\bibinfo {author} {\bibfnamefont {A.~P.}\ \bibnamefont
			{Solon}}\ and\ \bibinfo {author} {\bibfnamefont {J.~M.}\ \bibnamefont
			{Horowitz}},\ }\bibfield  {title} {\bibinfo {title} {{Phase Transition in
				Protocols Minimizing Work Fluctuations}},\ }\href
	{https://doi.org/10.1103/PhysRevLett.120.180605} {\bibfield  {journal}
		{\bibinfo  {journal} {Phys. Rev. Lett.}\ }\textbf {\bibinfo {volume} {120}},\
		\bibinfo {pages} {180605} (\bibinfo {year} {2018})}\BibitemShut {NoStop}%
	\bibitem [{\citenamefont {Seoane}\ and\ \citenamefont
		{Sol{\'e}}(2016)}]{Seoane2016}%
	\BibitemOpen
	\bibfield  {author} {\bibinfo {author} {\bibfnamefont {L.~F.}\ \bibnamefont
			{Seoane}}\ and\ \bibinfo {author} {\bibfnamefont {R.}~\bibnamefont
			{Sol{\'e}}},\ }\bibfield  {title} {\bibinfo {title} {{Multiobjective
				Optimization and Phase Transitions}},\ }in\ \href@noop {} {\emph {\bibinfo
			{booktitle} {Proceedings of ECCS 2014}}},\ \bibinfo {editor} {edited by\
		\bibinfo {editor} {\bibfnamefont {S.}~\bibnamefont {Battiston}}, \bibinfo
		{editor} {\bibfnamefont {F.}~\bibnamefont {De~Pellegrini}}, \bibinfo {editor}
		{\bibfnamefont {G.}~\bibnamefont {Caldarelli}},\ and\ \bibinfo {editor}
		{\bibfnamefont {E.}~\bibnamefont {Merelli}}}\ (\bibinfo  {publisher}
	{Springer International Publishing},\ \bibinfo {address} {Cham},\ \bibinfo
	{year} {2016})\ pp.\ \bibinfo {pages} {259--270}\BibitemShut {NoStop}%
	\bibitem [{Note3()}]{Note3}%
	\BibitemOpen
	\bibinfo {note} {Here, we have used that at steady state $\ev {\left
			(r_{k+1}\right )^{2}} = \ev {\left (r_{k}\right )^{2}}$.}\BibitemShut {Stop}%
	\bibitem [{\citenamefont {Wales}\ and\ \citenamefont {Doye}(1997)}]{Wales1997}%
	\BibitemOpen
	\bibfield  {author} {\bibinfo {author} {\bibfnamefont {D.~J.}\ \bibnamefont
			{Wales}}\ and\ \bibinfo {author} {\bibfnamefont {J.~P.~K.}\ \bibnamefont
			{Doye}},\ }\bibfield  {title} {\bibinfo {title} {{Global Optimization by
				Basin-Hopping and the Lowest Energy Structures of Lennard-Jones Clusters
				Containing up to 110 Atoms}},\ }\href {https://doi.org/10.1021/jp970984n}
	{\bibfield  {journal} {\bibinfo  {journal} {J. Phys. Chem. A}\ }\textbf
		{\bibinfo {volume} {101}},\ \bibinfo {pages} {5111} (\bibinfo {year}
		{1997})}\BibitemShut {NoStop}%
	\bibitem [{\citenamefont {Kim}\ and\ \citenamefont {Qian}(2007)}]{Kim2007}%
	\BibitemOpen
	\bibfield  {author} {\bibinfo {author} {\bibfnamefont {K.~H.}\ \bibnamefont
			{Kim}}\ and\ \bibinfo {author} {\bibfnamefont {H.}~\bibnamefont {Qian}},\
	}\bibfield  {title} {\bibinfo {title} {Fluctuation theorems for a molecular
			refrigerator},\ }\href {https://doi.org/10.1103/PhysRevE.75.022102}
	{\bibfield  {journal} {\bibinfo  {journal} {Phys. Rev. E}\ }\textbf {\bibinfo
			{volume} {75}},\ \bibinfo {pages} {022102} (\bibinfo {year}
		{2007})}\BibitemShut {NoStop}%
	\bibitem [{\citenamefont {Horowitz}\ and\ \citenamefont
		{Sandberg}(2014)}]{Horowitz2014}%
	\BibitemOpen
	\bibfield  {author} {\bibinfo {author} {\bibfnamefont {J.~M.}\ \bibnamefont
			{Horowitz}}\ and\ \bibinfo {author} {\bibfnamefont {H.}~\bibnamefont
			{Sandberg}},\ }\bibfield  {title} {\bibinfo {title} {Second-law-like
			inequalities with information and their interpretations},\ }\href
	{https://doi.org/10.1088/1367-2630/16/12/125007} {\bibfield  {journal}
		{\bibinfo  {journal} {New J. Phys.}\ }\textbf {\bibinfo {volume} {16}},\
		\bibinfo {pages} {125007} (\bibinfo {year} {2014})}\BibitemShut {NoStop}%
	\bibitem [{\citenamefont {H\"anggi}\ \emph {et~al.}(1990)\citenamefont
		{H\"anggi}, \citenamefont {Talkner},\ and\ \citenamefont
		{Borkovec}}]{Hanggi1990}%
	\BibitemOpen
	\bibfield  {author} {\bibinfo {author} {\bibfnamefont {P.}~\bibnamefont
			{H\"anggi}}, \bibinfo {author} {\bibfnamefont {P.}~\bibnamefont {Talkner}},\
		and\ \bibinfo {author} {\bibfnamefont {M.}~\bibnamefont {Borkovec}},\
	}\bibfield  {title} {\bibinfo {title} {{Reaction-rate theory: fifty years
				after Kramers}},\ }\href {https://doi.org/10.1103/RevModPhys.62.251}
	{\bibfield  {journal} {\bibinfo  {journal} {Rev. Mod. Phys.}\ }\textbf
		{\bibinfo {volume} {62}},\ \bibinfo {pages} {251} (\bibinfo {year}
		{1990})}\BibitemShut {NoStop}%
	\bibitem [{\citenamefont {Pontryagin}\ \emph {et~al.}(1933)\citenamefont
		{Pontryagin}, \citenamefont {Andronov},\ and\ \citenamefont
		{Vitt}}]{Pontryagin1933}%
	\BibitemOpen
	\bibfield  {author} {\bibinfo {author} {\bibfnamefont {L.}~\bibnamefont
			{Pontryagin}}, \bibinfo {author} {\bibfnamefont {A.}~\bibnamefont
			{Andronov}},\ and\ \bibinfo {author} {\bibfnamefont {A.}~\bibnamefont
			{Vitt}},\ }\bibfield  {title} {\bibinfo {title} {On the statistical treatment
			of dynamical systems},\ }\href@noop {} {\bibfield  {journal} {\bibinfo
			{journal} {Zh. Eksp. Teor. Fiz.}\ }\textbf {\bibinfo {volume} {3}},\ \bibinfo
		{pages} {165} (\bibinfo {year} {1933})},\ \bibinfo {note} {[English
		translation in ``Noise in. Nonlinear Dynamical Systems'', ed. by F. Moss,
		P.V.E. McClintock, Cambridge Univ. Press, Cambridge, vol. 1, p.
		329-348]}\BibitemShut {NoStop}%
	\bibitem [{\citenamefont {Park}\ \emph {et~al.}(2016)\citenamefont {Park},
		\citenamefont {Lee},\ and\ \citenamefont {Noh}}]{Park2016}%
	\BibitemOpen
	\bibfield  {author} {\bibinfo {author} {\bibfnamefont {J.-M.}\ \bibnamefont
			{Park}}, \bibinfo {author} {\bibfnamefont {J.~S.}\ \bibnamefont {Lee}},\ and\
		\bibinfo {author} {\bibfnamefont {J.~D.}\ \bibnamefont {Noh}},\ }\bibfield
	{title} {\bibinfo {title} {{Optimal tuning of a confined Brownian information
				engine}},\ }\href {https://doi.org/10.1103/PhysRevE.93.032146} {\bibfield
		{journal} {\bibinfo  {journal} {Phys. Rev. E}\ }\textbf {\bibinfo {volume}
			{93}},\ \bibinfo {pages} {032146} (\bibinfo {year} {2016})}\BibitemShut
	{NoStop}%
	\bibitem [{Note4()}]{Note4}%
	\BibitemOpen
	\bibinfo {note} {We construct the ratchet-time distribution by computing, for
		a given trajectory, the time $t_{\protect \mathrm {ratchet}}$ between
		trap-center movements along that trajectory. We histogram these times
		(properly normalized) to obtain the densities in Fig.~\ref
		{fig:ratchet_time_distr}.}\BibitemShut {Stop}%
	\bibitem [{\citenamefont {Bechhoefer}(2021)}]{bechhoefer_book2021}%
	\BibitemOpen
	\bibfield  {author} {\bibinfo {author} {\bibfnamefont {J.}~\bibnamefont
			{Bechhoefer}},\ }\href@noop {} {\emph {\bibinfo {title} {Control Theory for
				Physicists}}}\ (\bibinfo  {publisher} {Cambridge University Press},\ \bibinfo
	{year} {2021})\BibitemShut {NoStop}%
	\bibitem [{\citenamefont {Schmitt}\ \emph {et~al.}(2015)\citenamefont
		{Schmitt}, \citenamefont {Parrondo}, \citenamefont {Linke},\ and\
		\citenamefont {Johansson}}]{schmitt2015}%
	\BibitemOpen
	\bibfield  {author} {\bibinfo {author} {\bibfnamefont {R.~K.}\ \bibnamefont
			{Schmitt}}, \bibinfo {author} {\bibfnamefont {J.~M.~R.}\ \bibnamefont
			{Parrondo}}, \bibinfo {author} {\bibfnamefont {H.}~\bibnamefont {Linke}},\
		and\ \bibinfo {author} {\bibfnamefont {J.}~\bibnamefont {Johansson}},\
	}\bibfield  {title} {\bibinfo {title} {Molecular motor efficiency is
			maximized in the presence of both power-stroke and rectification through
			feedback},\ }\href {https://doi.org/10.1088/1367-2630/17/6/065011} {\bibfield
		{journal} {\bibinfo  {journal} {New J. Phys.}\ }\textbf {\bibinfo {volume}
			{17}},\ \bibinfo {pages} {065011} (\bibinfo {year} {2015})}\BibitemShut
	{NoStop}%
	\bibitem [{\citenamefont {Chang}\ \emph {et~al.}(2021)\citenamefont {Chang},
		\citenamefont {Chiang}, \citenamefont {Jun}, \citenamefont {Lai},\ and\
		\citenamefont {Chen}}]{Chang2021}%
	\BibitemOpen
	\bibfield  {author} {\bibinfo {author} {\bibfnamefont {H.}~\bibnamefont
			{Chang}}, \bibinfo {author} {\bibfnamefont {K.-H.}\ \bibnamefont {Chiang}},
		\bibinfo {author} {\bibfnamefont {Y.}~\bibnamefont {Jun}}, \bibinfo {author}
		{\bibfnamefont {P.-Y.}\ \bibnamefont {Lai}},\ and\ \bibinfo {author}
		{\bibfnamefont {Y.-F.}\ \bibnamefont {Chen}},\ }\bibfield  {title} {\bibinfo
		{title} {Generation of virtual potentials by controlled feedback in electric
			circuit systems},\ }\href {https://doi.org/10.1103/PhysRevE.103.042138}
	{\bibfield  {journal} {\bibinfo  {journal} {Phys. Rev. E}\ }\textbf {\bibinfo
			{volume} {103}},\ \bibinfo {pages} {042138} (\bibinfo {year}
		{2021})}\BibitemShut {NoStop}%
\end{thebibliography}
\end{document}